\documentclass[prd,nofootinbib,preprint,superscriptaddress]{revtex4}
\pdfoutput=1
\usepackage{amsmath,amssymb}
\usepackage{epsfig}
\usepackage{graphicx}
\usepackage[usenames,dvipsnames]{color}
\usepackage{subfigure}
\usepackage{slashed}
\usepackage{comment}
\usepackage[normalem]{ulem}
\usepackage[colorlinks,citecolor=blue]{hyperref}
\usepackage{color}

\newcommand{\bea}{\begin{eqnarray}}
\newcommand{\eea}{\end{eqnarray}}
\usepackage{multirow}

\usepackage{tikz-feynman}
\tikzfeynmanset{compat=1.1.0}

\usepackage{tikzsymbols}

\usepackage{xcolor}
\usepackage{soul}

\begin{document}

\title{Large neutrino asymmetry from forbidden decay of dark matter}

\author{Debasish Borah}
\email{dborah@iitg.ac.in}
\affiliation{Department of Physics, Indian Institute of Technology Guwahati, Assam 781039, India}
\author{Nayan Das}
\email{nayan.das@iitg.ac.in}
\affiliation{Department of Physics, Indian Institute of Technology
Guwahati, Assam 781039, India}
\author{Indrajit Saha}
\email{s.indrajit@iitg.ac.in}
\affiliation{Department of Physics, Indian Institute of Technology Guwahati, Assam 781039, India}

\begin{abstract}
Dark matter (DM), in spite of being stable or long-lived on cosmological scales, can decay in the early Universe due to finite-temperature effects. In particular, a first order phase transition (FOPT) in the early Universe can provide a finite window for such decay, guaranteeing DM stability at lower temperatures, consistent with observations. The FOPT can lead to the generation of stochastic gravitational waves (GW) with peak frequencies correlated with DM mass. On the other hand, early DM decay into neutrinos can create a large neutrino asymmetry which can have interesting cosmological consequences in terms of enhanced effective relativistic degrees of freedom $N_{\rm eff}$, providing a solution to the recently observed Helium anomaly among others. Allowing DM decay to occur below sphaleron decoupling temperature, thereby avoiding overproduction of baryon asymmetry, forces the FOPT to occur at sub-electroweak scale. This leaves the stochastic GW within range of experiments like LISA, $\mu$ARES, NANOGrav etc. 
\end{abstract}

\maketitle 

\section{Introduction}\label{sec:intro}
As suggested by several astrophysical and cosmological observations \cite{Zyla:2020zbs, Aghanim:2018eyx}, approximately $26\%$ of the present Universe is made up of DM, the particle origin of which is not yet known. Similar observations also suggest that the baryon content of the Universe is highly asymmetric leading to the longstanding puzzle of baryon asymmetry of Universe (BAU). This observed excess of baryons over anti-baryons is quantified in terms of the baryon to photon ratio as \cite{Planck:2018vyg} 
\begin{equation}
\eta_B = \frac{n_{B}-n_{\overline{B}}}{n_{\gamma}} \simeq 6.2 \times 10^{-10}, 
\label{etaBobs}
\end{equation} 
based on the cosmic microwave background (CMB) measurements which also agrees well with the big bang nucleosynthesis (BBN) estimates \cite{Zyla:2020zbs}.
While the weakly interacting massive particle (WIMP) \cite{Kolb:1990vq, Jungman:1995df, Bertone:2004pz, Feng:2010gw, Arcadi:2017kky, Roszkowski:2017nbc} has been the most widely studied particle DM candidate, baryogenesis/leptogenesis \cite{Weinberg:1979bt, Kolb:1979qa, Fukugita:1986hr} are considered to be the popular explanation for the BAU. However, none of the WIMP candidates nor baryogenesis/leptogenesis framework has been verified experimentally. While WIMP can have observable scattering rate with nucleons, persistent null results at direct detection experiments \cite{LZ:2022lsv,LZ:2024} have pushed WIMP DM to a tight corner. On the other hand, baryogenesis and leptogenesis typically remain a high scale phenomena with limited or no direct experimental detection prospects. However, depending upon the particular model, such scenarios can have promising indirect detection prospects. One such avenue is the detection of stochastic gravitational wave (GW) background, which has been utilised in several baryogenesis or leptogenesis scenarios \cite{Hall:2019ank, Dror:2019syi, Blasi:2020wpy, Fornal:2020esl, Samanta:2020cdk, Barman:2022yos, Baldes:2021vyz, Azatov:2021irb, Huang:2022vkf, Dasgupta:2022isg, Barman:2022pdo, Datta:2022tab, Borah:2022cdx, Borah:2023saq, Borah:2023god, Barman:2023fad} as well as particle DM models \cite{Hall:2019rld, Yuan:2021ebu, Tsukada:2020lgt, Chatrchyan:2020pzh, Bian:2021vmi, Samanta:2021mdm, Borah:2022byb, Azatov:2021ifm, Azatov:2022tii, Baldes:2022oev, Borah:2022iym, Borah:2022vsu, Shibuya:2022xkj, Borah:2023god, Borah:2023saq, Borah:2023sbc, Borah:2024lml, Adhikary:2024btd}. Another promising avenue is the signatures at CMB due to enhanced effective relativistic degrees of freedom $N_{\rm eff}$ in certain leptogenesis and DM models like \cite{FileviezPerez:2019cyn, Nanda:2019nqy, Han:2020oet, Mahanta:2021plx, Biswas:2021kio, Borah:2024gql}.

If a mechanism to explain BAU also leaves a large neutrino asymmetry at low temperatures, it can have interesting implications for observations related to BBN and CMB. Although the observed BAU restricts the net lepton asymmetry around sphaleron decoupling temperature $(T_{\rm sph})$ to be of same order as $\eta_B$, it is possible to have a large lepton asymmetry at post-sphaleron epoch $(T < T_{\rm sph})$ if the asymmetry is stored in neutrinos\footnote{Charge neutrality of the early Universe restricts the charge lepton asymmetry to be $\leq \eta_B$.}. Such large neutrino asymmetry can enhance the effective relativistic degrees of freedom $N_{\rm eff}$ which can be measured at CMB experiments \cite{Escudero:2022okz}. Additionally, a large neutrino asymmetry can also alter the predictions of BBN with observable consequences \cite{Mangano:2010ei, Mangano:2011ip, Castorina:2012md, Escudero:2022okz, Froustey:2024mgf}. A large neutrino asymmetry also allows resonant production of sterile neutrino DM via Shi-Fuller mechanism \cite{Shi:1998km} \footnote{One may refer to a review \cite{Drewes:2016upu} for details of these mechanisms and relevant experimental constraints.}. Alternately, such asymmetries can also alter the cosmological bounds on light sterile neutrinos \cite{Mirizzi:2012we}. Large neutrino asymmetry can also have other consequences for gravitational wave by turning the QCD phase transition to first order \cite{Schwarz:2009ii}, baryogenesis, \cite{Barrie:2017mmr, Gao:2024fhm}, astrophysical structure formation \cite{Zeng:2018pcv} etc. 

Motivated by this, we propose a new mechanism to create such large lepton asymmetry at post-sphaleron epoch in order not to be in conflict with the observed baryon asymmetry. While it is possible to create large lepton asymmetries in individual flavours at high scale while keeping the net lepton asymmetry $\sim \mathcal{O}(\eta_B)$ \cite{March-Russell:1999hpw}, there exist tight constraints on such setup due to the generation of helical hypermagnetic field which can source a new contribution to the baryon asymmetry of the Universe, as pointed out recently in \cite{Domcke:2022uue}. Among the scenarios considered so far to generate large neutrino asymmetry, one recent attempt \cite{Kawasaki:2022hvx} along with a few related earlier works \cite{March-Russell:1999hpw, McDonald:1999in, Casas:1997gx} have utilised the Affleck-Dine mechanism \cite{Affleck:1984fy}. The authors of \cite{Borah:2022uos} considered TeV scale leptogenesis from decay and scatterings as source of such lepton asymmetry generated at low scales while also creating the required lepton asymmetry at $T = T_{\rm sph}$ to generate the observed BAU. In \cite{ChoeJo:2023cnx}, a two-phase leptogenesis model was proposed where sub-electroweak scale right handed neutrino (RHN) can generate lepton asymmetries at two different scales due to finite-temperature effects on its mass. In this work, we consider the possibility of generating such large lepton asymmetry from forbidden decay of DM. Although DM is stable on cosmological scales, it can decay in the early Universe due to finite-temperature effects. Role of such forbidden DM decay on generating BAU was discussed in earlier works \cite{Borah:2023qag, Borah:2023god}. We consider initially overproduced asymmetric DM \cite{Nussinov:1985xr, Davoudiasl:2012uw, Petraki:2013wwa, Zurek:2013wia,DuttaBanik:2020vfr, Barman:2021ost, Cui:2020dly} to be the source of such large neutrino asymmetry. The decay of DM into neutrino is facilitated by a first order phase transition (FOPT) providing a finite window in which DM can decay, while ensuring its stability at later epochs. In addition to the cosmological consequences of such large neutrino asymmetry, the FOPT at post-sphaleron epoch leads to stochastic GW in nHz-mHz ballpark capable of explaining the recent observations of pulsar timing array (PTA) like NANOGrav \cite{NANOGrav:2023gor} as well.

This paper is organised as follows. In section \ref{sec1}, we discuss the setup to implement the idea. In section \ref{sec2}, we discuss the generation of large neutrino asymmetry from forbidden decay of DM followed by brief discussion of $N_{\rm eff}$ due to large neutrino asymmetry in section \ref{sec2a}. In section \ref{sec3}, we discuss our results followed by brief remarks on possible UV completion of our basic setup in section \ref{sec4}. We finally conclude in section \ref{sec5}.

\section{The framework}
\label{sec1}
We first consider a minimal setup consisting of a Dirac singlet fermion $\chi$, two singlet scalars $\Phi_{1,2}$ as the new degrees of freedom beyond the standard model (BSM). An unbroken $Z_2$ symmetry under which $\chi, \Phi_2$ are odd while all other fields are even ensures the stability of DM. The low energy effective Lagrangian is 
\begin{equation}
    -\mathcal{L} \supset \big ( y_1 \overline{\chi} \Phi_2 \nu + {\rm h.c.} \big ) + m_0 \bar{\chi}\chi + y_{\chi\Phi_1}\bar{\chi}\Phi_1\chi + V(\Phi_1, \Phi_2)
    \label{eq2}
\end{equation}
where $V(\Phi_1, \Phi_2)$, the tree level scalar potential involving the singlet scalars, is given by
\begin{align}
   V_{\rm tree} \equiv V(\Phi_1,\Phi_2) & =\frac{\lambda_1}{4} \left (\Phi_1^2-\frac{v_D^2}{2} \right)^2 + \mu_{\Phi_2}^2|\Phi_2|^2 + \lambda_{2}|\Phi_2|^4 + \frac{\lambda_{\Phi_1\Phi_2}}{2}\Phi_1^2 |\Phi_2|^2 
    \label{Vtree}
\end{align}
In the above scalar potential $v_D$ is the vacuum expectation value (VEV) acquired by the real scalar field $\Phi_1$ while driving the FOPT. It is worth mentioning that terms with odd powers of $\Phi_1$ can also be there in the above scalar potential as they are allowed by the chosen symmetries of the model. We have ignored them for simplicity as it keeps the number of free parameters minimal. The terms considered in the scalar potential of Eq. \eqref{Vtree} are sufficient to induce a first-order phase transition and ensure forbidden decay of dark matter $\chi$ due to sudden change of $\Phi_2$ mass at nucleation temperature, the details of which are discussed below. A dark global $U(1)_D$ symmetry is considered under which $\Phi_1$ transforms trivially whereas $\chi, 
\Phi_2$ have $U(1)_D$ charges of $1$ each. 

We calculate the complete potential including the tree level potential $V_{\rm tree}$, one-loop Coleman-Weinberg potential $V_{\rm CW}$ \cite{Coleman:1973jx} along with the finite-temperature potential $V_{\rm th}$ \cite{Dolan:1973qd,Quiros:1999jp}. The thermal field-dependent masses of different particles coupled to the singlet scalar $\Phi_1$ are incorporated in the full potential. The zero-temperature masses of DM and $\Phi_2$ are denoted by $m_\chi$ and $m_{\Phi_2}$ respectively. We then calculate the critical temperature $T_c$ at which the potential in the $\Phi_1$ direction acquires another degenerate minima at $v_c = \phi_1 (T=T_c)$. The order parameter of the FOPT is defined as $v_c/T_c$ such that a stronger FOPT corresponds to a larger $v_c/T_c$. The FOPT then proceeds via tunneling, the rate of which is estimated by calculating the bounce action $S_3$ using the prescription in \cite{Linde:1980tt, Adams:1993zs}. The nucleation temperature $T_n$ is then calculated by comparing the tunneling rate with the Hubble expansion rate of the Universe $\Gamma (T_n) = \mathcal{H}^4(T_n)$. We also estimate the reheat temperature $T_{\rm RH}$ at the end of the FOPT due to the release of radiation energy. It is defined as $T_{\rm RH} = {\rm Max}[T_n, T_{\rm inf}]$ \cite{Baldes:2021vyz} where $T_{\rm inf}$ is determined by equating density of radiation energy to that of energy released from the FOPT or equivalently $\Delta V_{\rm eff}$. The details of the FOPT is given in Appendix \ref{appen1}.

\begin{figure}
    \centering
    \includegraphics[width=0.75\linewidth]{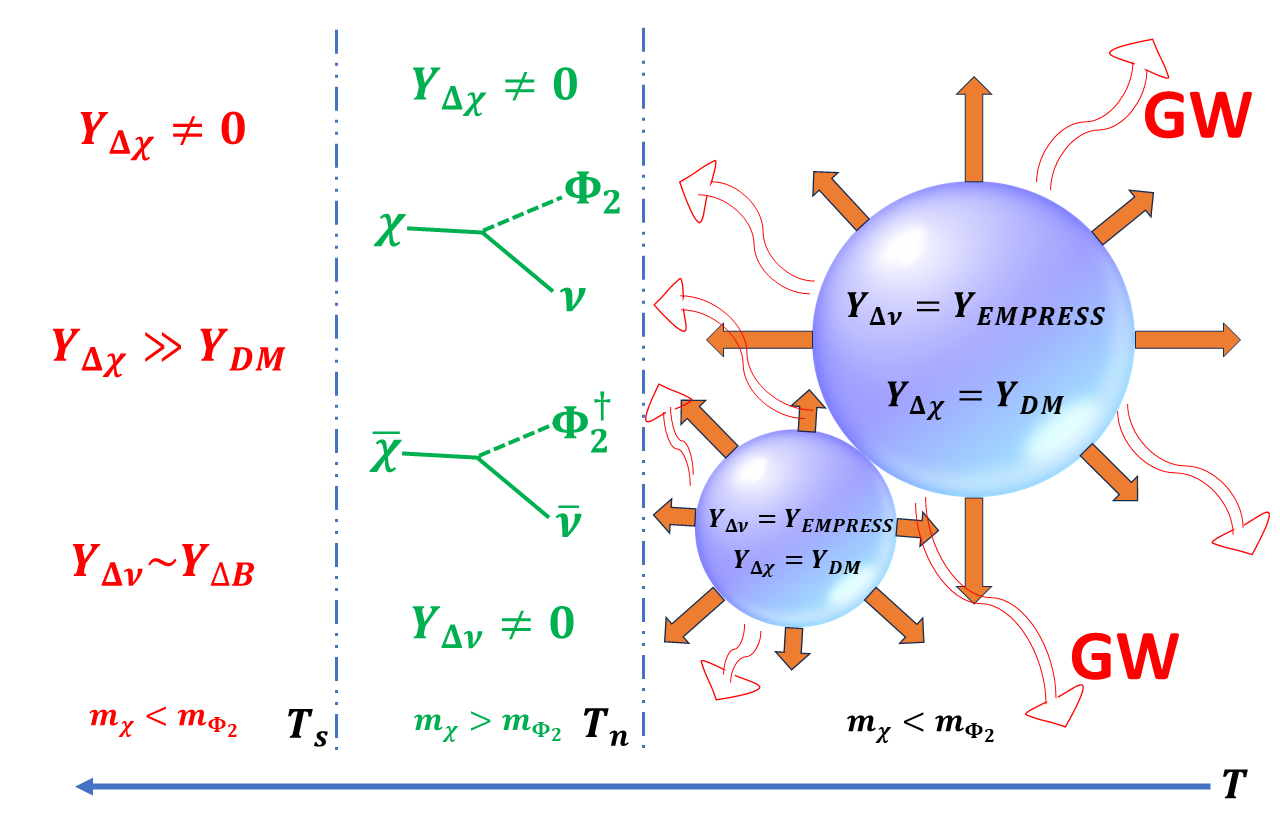}
    \caption{The schematic diagram of co-genesis}
    \label{fig:schematic}
\end{figure}

We consider a large initial dark sector asymmetry $Y_{\Delta \chi} \neq 0$ while lepton asymmetry being of the same order as baryon asymmetry to be consistent with the observed BAU. At a temperature $T=T_s$ below the sphaleron decoupling $T_{\rm sph}$, the decay of $\chi$ into $\Phi_2 \, \nu$ gets allowed enabling the transfer of dark sector asymmetry into neutrino asymmetry $Y_{\Delta \nu}$. At the nucleation temperature $T=T_n$, the decay $\chi \rightarrow \Phi_2 \, \nu$ gets forbidden again due to sharp increase in $\Phi_2$ mass. This freezes the production of neutrino asymmetry while ensuring the stability of DM at lower temperatures. Fig. \ref{fig:schematic} represents a schematic of this idea.

\section{Co-genesis of Large neutrino asymmetry and Dark Matter}
\label{sec2}
In order to calculate the final neutrino asymmetry and DM abundance starting with an initial dark asymmetry, we write the relevant Boltzmann equations for $\chi, \Phi_2, \nu$ and their anti-particles $\bar{\chi}, \Phi_2^\dagger, \bar{\nu}$. We consider temperature dependent masses of relevant particles such that $\chi, \bar{\chi}$ decay is allowed only for a finite window $T_s > T > T_n$. While neutrinos do not receive much thermal corrections below the electroweak scale, the thermal mass of $\Phi_2$ plays crucial role in determining the final asymmetries. The relevant Boltzmann equations for generating a large neutrino asymmetry from an initial dark sector asymmetry can be written as follows.
\begin{align}
    \frac{d Y_{\chi}}{dz} & =-\frac{1}{\mathcal{H}z}\langle\Gamma_{\chi}\rangle\Theta(m_{\chi}-m_{\Phi_2}) \left(Y_{\chi}-Y_{\chi}^{\rm eq}\frac{Y_{\Phi_2}}{Y_{\Phi_2}^{\rm eq}}\right) + \frac{1}{\mathcal{H}z}\langle \Gamma_{\Phi_2} \rangle \Theta(m_{\Phi_2}-m_{\chi}) \left(Y_{\Phi_2}-Y_{\Phi_2}^{\rm eq}\frac{Y_{\chi}}{Y_{\chi}^{\rm eq}}\right) \nonumber \\
    &-\frac{s}{\mathcal{H}z}\langle\sigma v_{\chi \bar{\chi}\rightarrow \Phi_1 \Phi_1}\rangle\left(Y_{\chi}Y_{\bar{\chi}}  - Y_{\chi}^{\rm eq} Y_{\bar{\chi}}^{\rm eq} \right),
    \nonumber 
\end{align}
\begin{align}
    \frac{d Y_{\bar{\chi}}}{dz} & =-\frac{1}{\mathcal{H}z}\langle\Gamma_{\bar{\chi}}\rangle\Theta(m_{\chi}-m_{\Phi_2}) \left(Y_{\bar{\chi}}-Y_{\bar{\chi}}^{\rm eq}\frac{Y_{\Phi_2^\dagger}}{Y_{\Phi_2^\dagger}^{\rm eq}}\right) + \frac{1}{\mathcal{H}z}\langle\Gamma_{\Phi_2^\dagger}\rangle\Theta(m_{\Phi_2}-m_{\chi}) \left(Y_{\Phi_2^\dagger}-Y_{\Phi_2^\dagger}^{\rm eq}\frac{Y_{\bar{\chi}}}{Y_{\bar{\chi}}^{\rm eq}}\right) \nonumber \\
    &-\frac{s}{\mathcal{H}z}\langle\sigma v_{\chi \bar{\chi}\rightarrow \Phi_1 \Phi_1}\rangle\left(Y_{\chi}Y_{\bar{\chi}}  - Y_{\chi}^{\rm eq} Y_{\bar{\chi}}^{\rm eq} \right),
    \nonumber
\end{align}
\begin{align}
    \frac{d Y_{\Phi_2}}{dz} &=\frac{1}{\mathcal{H}z}\langle\Gamma_{\chi}\rangle\Theta(m_{\chi}-m_{\Phi_2}) \left(Y_{\chi}-Y_{\chi}^{\rm eq}\frac{Y_{\Phi_2}}{Y_{\Phi_2}^{\rm eq}}\right) - \frac{1}{\mathcal{H}z}\langle\Gamma_{\Phi_2}\rangle\Theta(m_{\Phi_2}-m_{\chi}) \left(Y_{\Phi_2}-Y_{\Phi_2}^{\rm eq}\frac{Y_{\chi}}{Y_{\chi}^{\rm eq}}\right) \nonumber \\
    &-\frac{s}{\mathcal{H}z}\langle\sigma v_{\Phi_2\Phi_2^\dagger\rightarrow \Phi_1\Phi_1}\rangle \left(Y_{\Phi_2}Y_{\Phi_2^\dagger}-Y_{\Phi_2}^{\rm eq}Y_{\Phi_2^\dagger}^{\rm eq}\right) -\frac{s}{\mathcal{H}z}\langle\sigma v_{\Phi_1\Phi_2\rightarrow \Phi_1\Phi_2^\dagger}\rangle Y_{\Phi_1}^{\rm eq}\left(Y_{\Phi_2}-Y_{\Phi_2^\dagger}\right),
    \nonumber 
\end{align}
\begin{align}
     \frac{d Y_{\Phi_2^\dagger}}{dz} &=\frac{1}{\mathcal{H}z}\langle\Gamma_{\bar{\chi}}\rangle\Theta(m_{\chi}-m_{\Phi_2}) \left(Y_{\bar{\chi}}-Y_{\bar{\chi}}^{\rm eq}\frac{Y_{\Phi_2^\dagger}}{Y_{\Phi_2^\dagger}^{\rm eq}}\right) - \frac{1}{\mathcal{H}z}\langle\Gamma_{\Phi_2^\dagger}\rangle\Theta(m_{\Phi_2}-m_{\chi}) \left(Y_{\Phi_2^\dagger}-Y_{\Phi_2^\dagger}^{\rm eq}\frac{Y_{\bar{\chi}}}{Y_{\bar{\chi}}^{\rm eq}}\right)\nonumber \\
    &-\frac{s}{\mathcal{H}z}\langle\sigma v_{\Phi_2\Phi_2^\dagger\rightarrow \Phi_1\Phi_1}\rangle \left(Y_{\Phi_2}Y_{\Phi_2^\dagger}-Y_{\Phi_2}^{\rm eq}Y_{\Phi_2^\dagger}^{\rm eq}\right)+\frac{s}{\mathcal{H}z}\langle\sigma v_{\Phi_1\Phi_2\rightarrow \Phi_1\Phi_2^\dagger}\rangle Y_{\Phi_1}^{\rm eq}\left(Y_{\Phi_2}-Y_{\Phi_2^\dagger}\right),
    \nonumber 
\end{align}
\begin{align}
    \frac{d Y_{ \nu }}{dz} =\frac{1}{\mathcal{H}z}\langle\Gamma_{\chi}\rangle\Theta(m_{\chi}-m_{\Phi_2}) \left(Y_{\chi}-Y_{\chi}^{\rm eq}\frac{Y_{\Phi_2}}{Y_{\Phi_2}^{\rm eq}}\right) + \frac{1}{\mathcal{H}z}\langle\Gamma_{\Phi_2^\dagger}\rangle\Theta(m_{\Phi_2}-m_{\chi}) \left(Y_{\Phi_2^\dagger}-Y_{\Phi_2^\dagger}^{\rm eq}\frac{Y_{\bar{\chi}}}{Y_{\bar{\chi}}^{\rm eq}}\right), \nonumber
\end{align}
\begin{align}
    \frac{d Y_{ \bar{\nu} }}{dz} =\frac{1}{\mathcal{H}z}\langle\Gamma_{\bar{\chi}}\rangle\Theta(m_{\chi}-m_{\Phi_2}) \left(Y_{\bar{\chi}}-Y_{\bar{\chi}}^{\rm eq}\frac{Y_{\Phi_2^\dagger}}{Y_{\Phi_2^\dagger}^{\rm eq}}\right)+ \frac{1}{\mathcal{H}z}\langle\Gamma_{\Phi_2}\rangle\Theta(m_{\Phi_2}-m_{\chi}) \left(Y_{\Phi_2}-Y_{\Phi_2}^{\rm eq}\frac{Y_{\chi}}{Y_{\chi}^{\rm eq}}\right).
\end{align}
Here $Y_i=n_i/s$ denotes comoving number density of species ``i" with $s$ being the entropy density. Hubble expansion rate is denoted by $\mathcal{H}$ while the variable $z$ is $m_\chi/T$. The thermal averaged decay rate \cite{Kolb:1979qa, Buchmuller:2004nz} for $\chi \rightarrow \Phi_2 \nu$ (and its conjugate) can be written as $\langle\Gamma_\chi\rangle=\frac{y^2_1}{32 \pi}m_\chi \left (1-\frac{m_{\Phi_2}^2}{m_\chi^2} \right)^2 \frac{K_1(m_\chi/T)}{K_2(m_\chi/T)}$ where $K_{1,2}$ are modified Bessel functions of 1st, 2nd kind respectively. Similarly, $\langle\Gamma_{\Phi_2}\rangle=\frac{y^2_1}{16 \pi}m_{\Phi_2}\left (1-\frac{m_{\chi}^2}{m_{\Phi_2}^2}\right )^2 \frac{K_1(m_{\Phi_2}/T)}{K_2(m_{\Phi_2}/T)}$. The $\Theta$ functions associated with these decay widths ensures the kinematical constraints. The equations for $\nu$ and $\bar{\nu}$ should also have collision terms corresponding to the SM gauge interactions on the right hand side. These interactions are responsible for keeping the neutrino number density close to their equilibrium values. However, we have ignored those terms as they do not contribute to the neutrino asymmetries. The symmetric part of $\chi$ can annihilate into a pair of light singlet scalars $\Phi_1$ with $\langle\sigma v_{\chi \bar{\chi}\rightarrow \Phi_1 \Phi_1}\rangle $ denoting the corresponding thermal-averaged cross-section \cite{Gondolo:1990dk}. A symmetry-breaking term of the form $\lambda_{12} (\Phi_2 \Phi_2 \Phi_1 \Phi_1 + \text{h.c.})$ is also introduced, ensuring that the asymmetry accumulated in the complex scalar $\Phi_2$ gets washed out. While other explicit $U(1)_D$ breaking terms like $\mu_{12} \Phi_2 \Phi^2_1 + \mu'_{12} \Phi^2_2 \Phi_1$ can also cause such washout of $\Phi_2$ asymmetry, we consider only the quartic term in numerical calculations for illustrative purpose.


Next we define $Y_{\Delta \chi}=Y_{\chi}-Y_{\bar{\chi}}$ and $Y_{\Delta \nu}=Y_{\nu}-Y_{\bar{\nu}}$. We choose the following initial condition for solving the above coupled  Boltzmann equations
\bea
Y_{\chi}(0)=Y_{\chi}^{\rm eq},~ Y_{\bar{\chi}}(0)=Y_{\chi}^{\rm eq}-Y^{\rm in}_{\Delta \chi},\\
Y_{\nu}(0)=Y_{\nu}^{\rm eq},~Y_{\bar{\nu}}(0)=Y_{\nu}^{\rm eq}.
\eea
The initial dark sector asymmetry $Y^{\rm in}_{\Delta \chi}$ in required amount can be generated in a variety of ways depending upon the UV completion discussed in section \ref{sec4}. Also, this initial DM asymmetry can be related to the origin of baryon asymmetry in the spirit of cogenesis within the framework of asymmetric dark matter (ADM) \cite{Nussinov:1985xr, Kaplan:2009ag} paradigm\footnote{See \cite{Petraki:2013wwa,Zurek:2013wia} for reviews of ADM scenarios.}.

\section{Large neutrino asymmetry and $N_{\rm eff}$}
\label{sec2a}
A large neutrino asymmetry can be parameterised in terms of chemical potential $\xi$ as follows \cite{Kawasaki:2022hvx},
\begin{equation}
   \eta_{\Delta L_\nu}= \frac{(n_\nu - n_{\overline{\nu}})}{n_\gamma}= \frac{\pi^2}{12 \zeta(3)} \sum_\alpha \xi_\alpha, \,\,\,\alpha \equiv e, \mu, \tau,
\end{equation}
\begin{align}
    Y_{\Delta \nu} = \sum_\alpha Y_{\Delta \nu_\alpha}=\sum_\alpha \frac{n_{\nu_\alpha} -n_{\bar{\nu}_\alpha}}{s} \simeq \sum_\alpha 0.035 \xi_\alpha.
\end{align}
Now, the presence of neutrino asymmetry alters the energy density of neutrinos which is incorporated to the parameter $N_{\rm eff}$, the effective number of relativistic species. Ignoring the effect of neutrino asymmetry on neutrino decoupling, $\Delta N_{\rm eff} = N_{\rm eff} - N^{\rm SM}_{\rm eff}$ is given as 
\begin{align} \label{eq:Neff1}
    \Delta N_{\rm eff}=\sum_\alpha \left[\frac{30}{7} \left(\frac{\xi_\alpha}{\pi} \right)^2 +\frac{15}{7} \left(\frac{\xi_\alpha}{\pi} \right)^4 \right].
\end{align}
Taking into account the effect of neutrino asymmetry on neutrino decoupling, the correction to the above expression can be written as \cite{Li:2024gzf}
\begin{align}\label{eq:Neff2}
    \Delta N_{\rm eff}=\sum_\alpha \left[\frac{30}{7} \left(\frac{\xi_\alpha}{\pi} \right)^2 +\frac{15}{7} \left(\frac{\xi_\alpha}{\pi} \right)^4 + \frac{0.0102}{3} \xi^2_{\alpha}\right].
\end{align}
As this is a small correction term, both Eq. \eqref{eq:Neff1} and Eq. \eqref{eq:Neff2} give the similar change in $N_{\rm eff}$ in presence of neutrino asymmetry.

In the SM, such neutrino asymmetries do not arise and hence there is no additional contribution to $N_{\rm eff}$. The sole contribution to $N_{\rm eff}$ in the SM comes from relativistic nature of light neutrinos. If neutrinos decouple instantaneously, then $N_{\rm eff}=3$. Considering non-instantaneous decoupling of SM neutrinos along with flavour oscillations and plasma correction of quantum electrodynamics, the SM $N_{\rm eff}$ value shifts to $N_{\rm eff}^{\rm SM} = 3.045$ \cite{Mangano:2005cc, Grohs:2015tfy,deSalas:2016ztq}\footnote{A few recent works \cite{Froustey:2020mcq, Bennett:2020zkv, Drewes:2024wbw} obtained a slightly different prediction $N_{\rm eff}^{\rm SM} = 3.044$.}. A deviation from $N_{\rm eff}^{\rm SM}$ indicates presence of BSM physics either in terms of additional light degrees of freedom, new interactions of SM neutrinos or large neutrino asymmetry mentioned above. The current bound on $N_{\rm eff}$ from Planck 2018 data is given as $N_{\rm eff}=2.99^{+0.34}_{-0.33}$ at $2\sigma$ CL including baryon acoustic oscillation (BAO) data. This corresponds to $\Delta N_{\rm eff} \lesssim 0.285$. The latest DESI 2024 data give a slightly weaker bound $\Delta N_{\rm eff} \lesssim 0.4$ at $2\sigma$ CL \cite{DESI:2024mwx}. Similar bound also exists from BBN considerations $2.3 < {\rm N}_{\rm eff} <3.4$ at $95\%$ CL \cite{Cyburt:2015mya}. Future CMB experiment CMB Stage IV (CMB-S4) is expected to reach a much better sensitivity of $\Delta {\rm N}_{\rm eff}={\rm N}_{\rm eff}-{\rm N}^{\rm SM}_{\rm eff}
= 0.06$ \cite{Abazajian:2019eic}, taking it closer to the SM prediction. Another future experiment CMB-HD \cite{CMB-HD:2022bsz} can probe $\Delta N_{\rm eff}$ upto $0.014$ at $1\sigma$.

In addition to enhancing $N_{\rm eff}$, a large neutrino asymmetry can also affect the BBN predictions of light nuclei abundance, as pointed out recently in the light of anomalous observations related to primordial Helium-4 ($^4{\rm He}$) abundance. The recent observations made by the Subaru Survey \cite{Matsumoto:2022tlr}, together with previous observations, have indicated a shift in primordial $^4{\rm He}$ abundance $Y_P=0.2379^{+0.0031}_{-0.0030}$ from the standard BBN predictions. While this is slightly smaller than earlier estimates \cite{2020ApJ, Aver:2015iza, Izotov:2014fga}, inclusion of the primordial deuterium constraints lead to a $>2\sigma$ tension between the predicted number of neutrino species $N_{\rm eff}=2.41^{+0.19}_{-0.21} $ and the SM predicted value $N_{\rm eff}=3.045$, referred to as the Helium anomaly \cite{Matsumoto:2022tlr}. With a large neutrino asymmetry in electron flavour $\xi_e = 0.05^{+0.03}_{-0.03}$, it is however possible to obtain a large $N_{\rm eff} = 3.22^{+0.33}_{-0.30}$ consistent with SM prediction within $1\sigma$ \cite{Matsumoto:2022tlr}. A combined analysis of BBN and CMB observations have also found evidence for a large neutrino asymmetry in the early Universe at $\sim 2\sigma$ confidence level \cite{Burns:2022hkq}.

\section{Results and Discussion}
\label{sec3}
\begin{figure}
    \centering
    \includegraphics[width=0.45\linewidth]{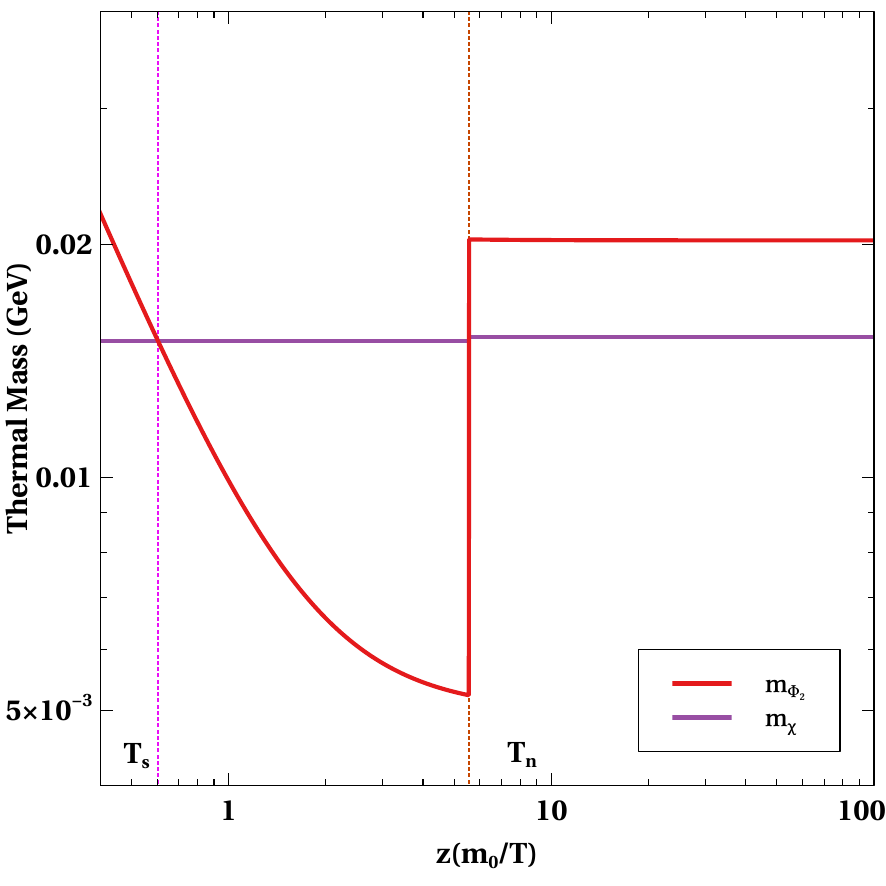}
      \includegraphics[width=0.45\linewidth]{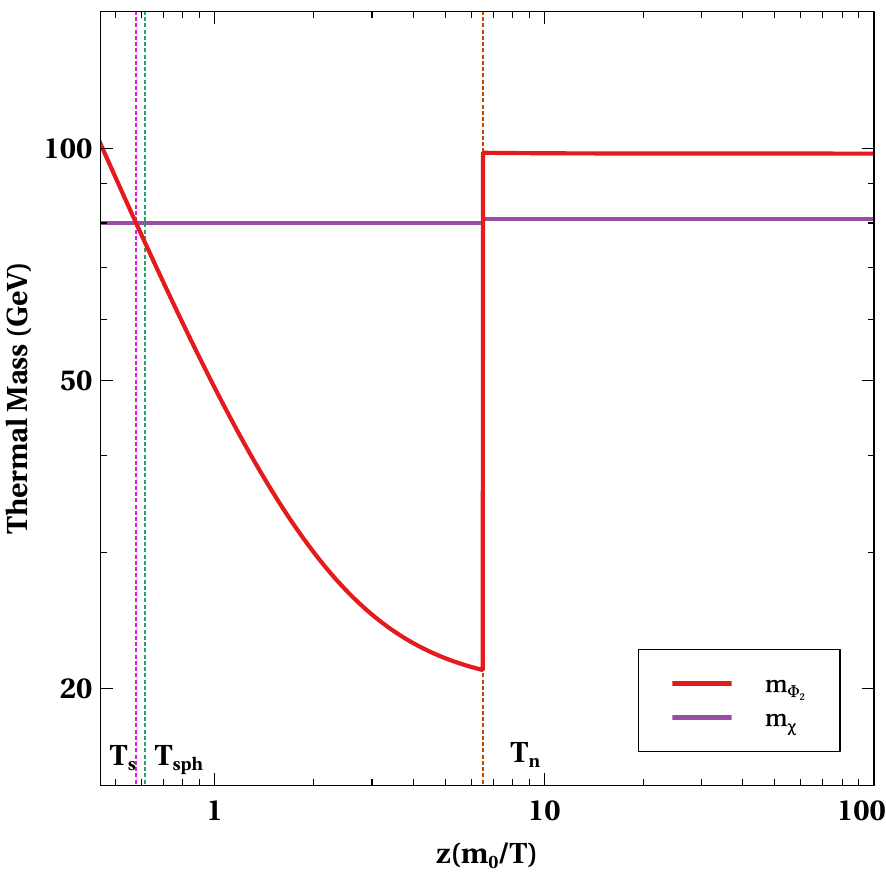} \\
     \includegraphics[width=0.45\linewidth]{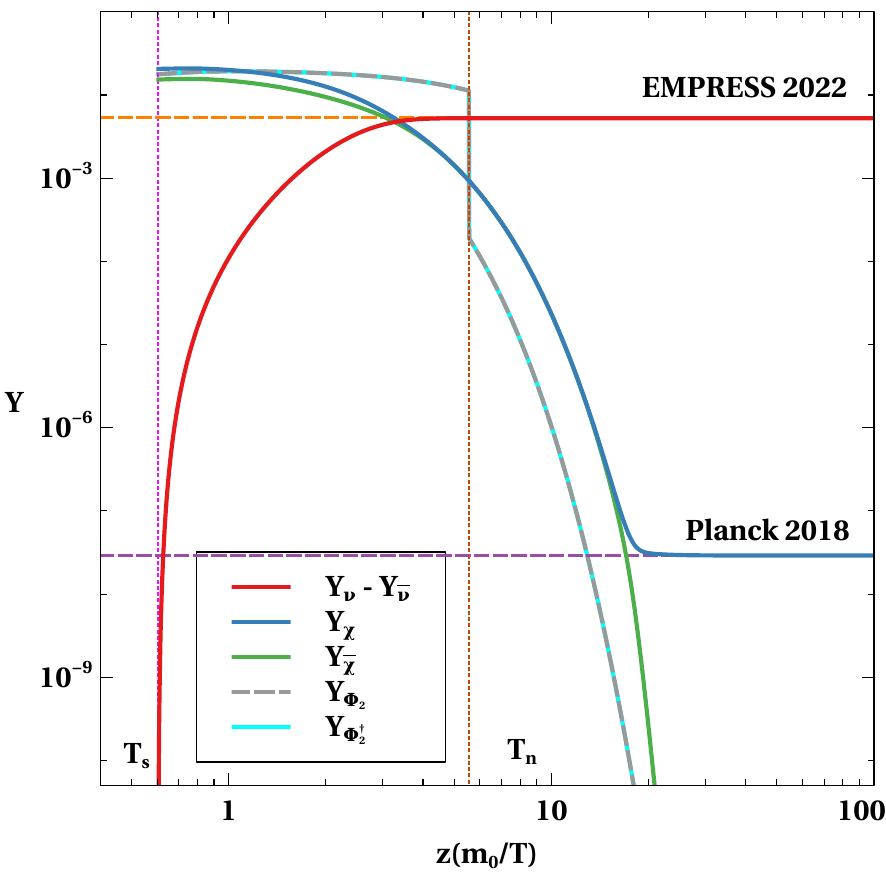}
          \includegraphics[width=0.45\linewidth]{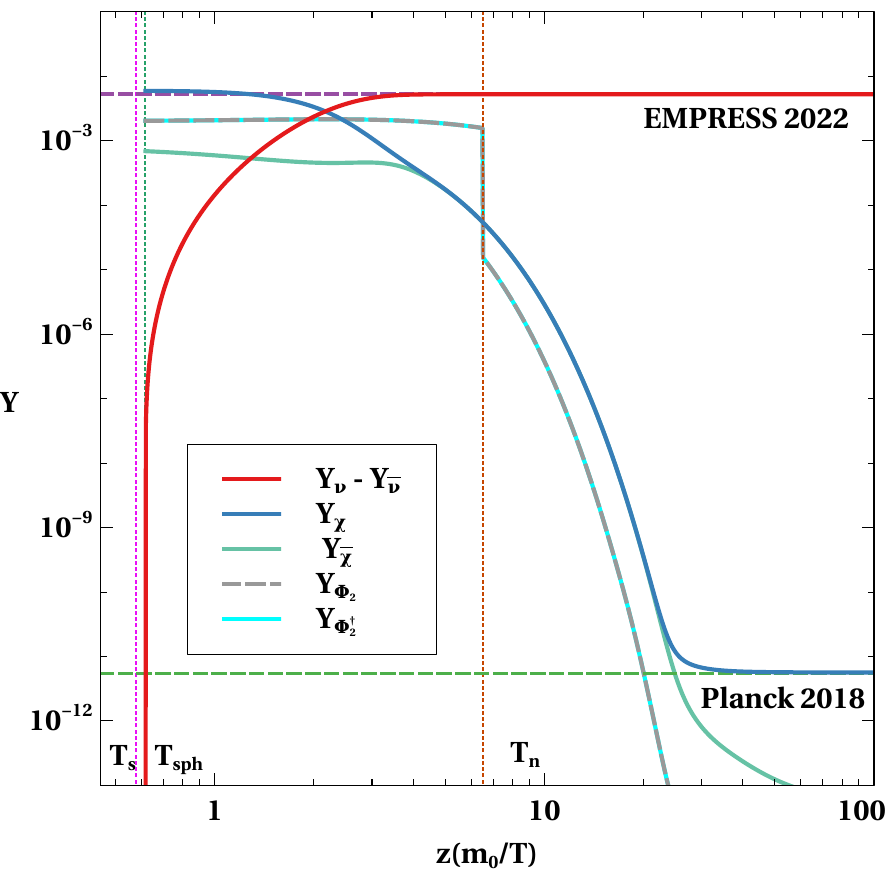}
   \caption{Top panel: Finite temperature masses of $\chi$ and $\Phi_2$ for BP1 (left) and BP6 (right) mentioned in table \ref{tab1}, \ref{tab2} with $m_{\chi}$=15.2$\times10^{-3}$ GeV (left) and $m_{\chi}$=81 GeV (right). Bottom panel: Evolution of comoving number densities for BP1 (left) and BP6 (right) with $y_1=7.34 \times10^{-10}$ (left) and $y_1=1.05\times10^{-7}$ (right).}
    \label{fig:thermal_mass1}
\end{figure}

We first discuss the temperature dependence of relevant particle masses leading to the finite temperature window allowing forbidden decay of DM. The top panel of Fig. \ref{fig:thermal_mass1} shows temperature dependence of masses of $\chi, \Phi_2$. While the details of scalar masses is given in Appendix \ref{appen1}, the thermal mass of $\chi$ \cite{Bellac:2011kqa,Laine:2016hma} is
\begin{equation}
    m_\chi^2(T)=m_\chi^2 +\frac{y^2_{\chi\Phi_1}}{16}T^2,
\end{equation}
with the correction being solely due to its Yukawa coupling. The top left and top right panel plots of Fig. \ref{fig:thermal_mass1} correspond to the benchmark points BP1, BP6 given in table \ref{tab1} and table \ref{tab2} respectively. Due to the small values of Yukawa coupling $y_{\chi \Phi_1}$ chosen, $\chi$ does not receive noticeable thermal corrections to its mass. Clearly, there exists a finite window $T_n < T < T_s$ during which $m_\chi > m_{\Phi_2}$ allowing the decay $\chi \rightarrow \Phi_2 \nu$ and its conjugate process. We also show the sphaleron decoupling temperature $T_{\rm sph} \sim 130$ GeV to indicate that the large neutrino asymmetry is produced only at $T < T_{\rm sph}$ and hence can not get converted into baryon asymmetry. The bottom panel plots of the same figure shows the corresponding evolution of comoving number densities of $\chi, \bar{\chi}, \Phi_2$, $\Phi^\dagger_2$ and $\Delta L_{\nu}$. Starting with an initial asymmetry $Y_\chi-Y_{\bar{\chi}} \neq 0$, the plots clearly depict the transfer of asymmetry in $\chi$ into a neutrino asymmetry during $T_n < T < T_s$. For $T < T_n$, the neutrino asymmetry saturates while $\chi, \bar{\chi}$ abundances continue to fall due to efficient annihilation $\chi \bar{\chi} \rightarrow \Phi_1 \Phi_1$. Due to the finite temperature window and choice of Yukawa coupling $y_1$ controlling the decay rate, all dark asymmetry does not get transferred to neutrinos, leaving a remnant asymmetry. The remnant asymmetric relic for the chosen benchmark points agree with the observed DM relic. Note that, $\Phi_2$ decay for $T < T_n$ can not alter asymmetries in dark sector or neutrinos as $\Phi_2$ does not store any asymmetry due to efficient $\Phi_2 \rightarrow \Phi^\dagger_2$ interactions. This is seen from the evolutions of $\Phi_2$ and $\Phi^\dagger_2$ which almost coincide with each other. The details of this conversion process can be found in Appendix \ref{appen3}.

Fig. \ref{fig:gw1} shows the GW spectra for the benchmark points given in tables \ref{tab1}, \ref{tab1a}, \ref{tab2} and \ref{tab2a}. Tables \ref{tab1a} and \ref{tab2a} contain the details of the GW related parameters used for calculating the spectra, while tables \ref{tab1}, \ref{tab2} contain the corresponding model parameters. In the total GW spectrum shown in Fig.~\ref{fig:gw1}, the sound-wave contribution dominates, with the peak corresponding to the sound-wave generated GW amplitude. This is the case for all benchmark points as the order of $\alpha_*$ and $\beta/\mathcal{H}_*$ are similar. The details of these GW sources during a FOPT can be found in appendix \ref{appen1}. Clearly, the peak frequencies can remain within the sensitivity of planned future experiment like LISA \cite{2017arXiv170200786A}, $\mu$ARES \cite{Sesana:2019vho}, THEIA~\cite{Garcia-Bellido:2021zgu}, GAIA \cite{Garcia-Bellido:2021zgu}, SKA~\cite{Weltman:2018zrl} and NANOGrav. The black colored violin-shaped points correspond to the recent NANOGrav data~\cite{NANOGrav:2023gor}. Due to the restrictions of having the FOPT below the electroweak scale, the peak frequencies do not come within reach of experiments like DECIGO \cite{Kawamura:2006up}, BBO \cite{Yagi:2011wg} among others.

\begin{figure}
    \centering
    \includegraphics[width=0.45\linewidth]{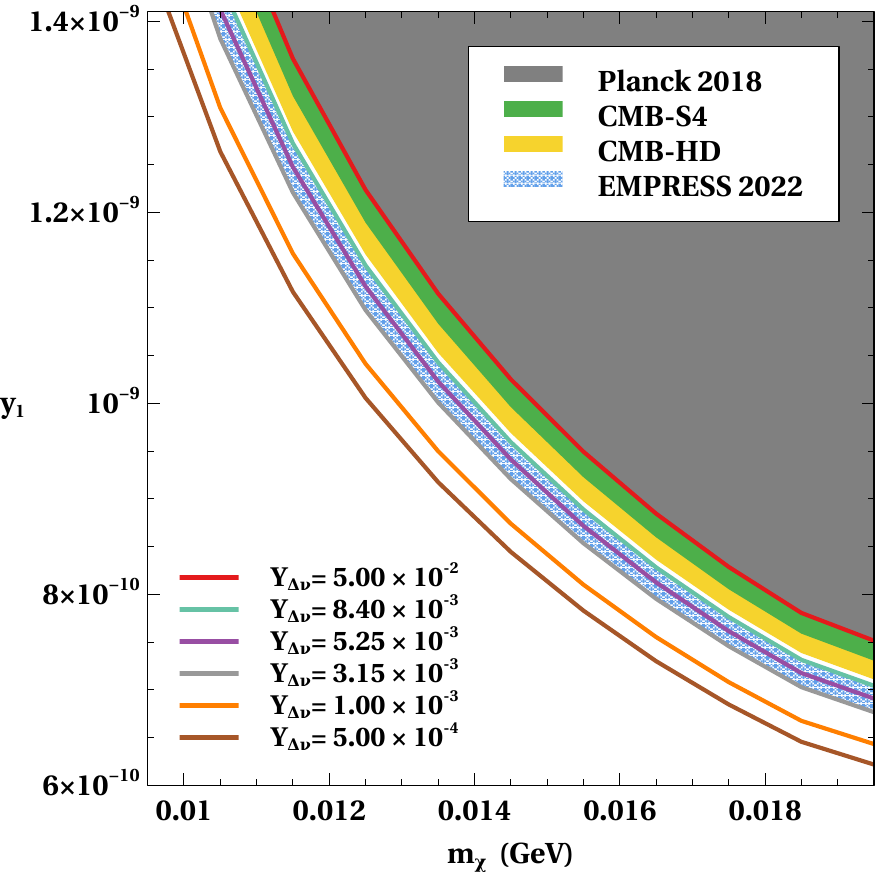}
        \includegraphics[width=0.45\linewidth]{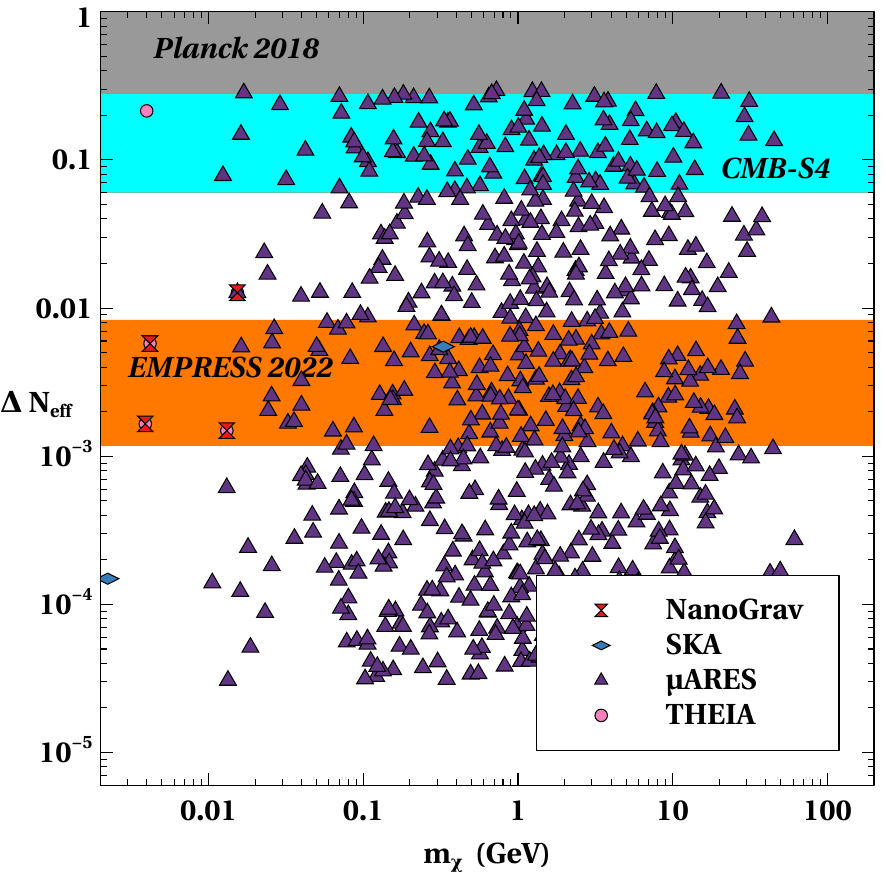}
    \caption{Left panel: Dark matter mass versus effective Yukawa coupling $y_1$ for different residual neutrino asymmetry in case of BP1 given in table \ref{tab1}. Right panel: Parameter space in $m_\chi$-$\Delta N_{\rm eff}$ plane with color code indicating SNR greater than 5 for future experiments and greater than 1 for ongoing experiments. }
    \label{summary1}
\end{figure}

\begin{figure}
    \centering
    \includegraphics[width=0.5\linewidth]{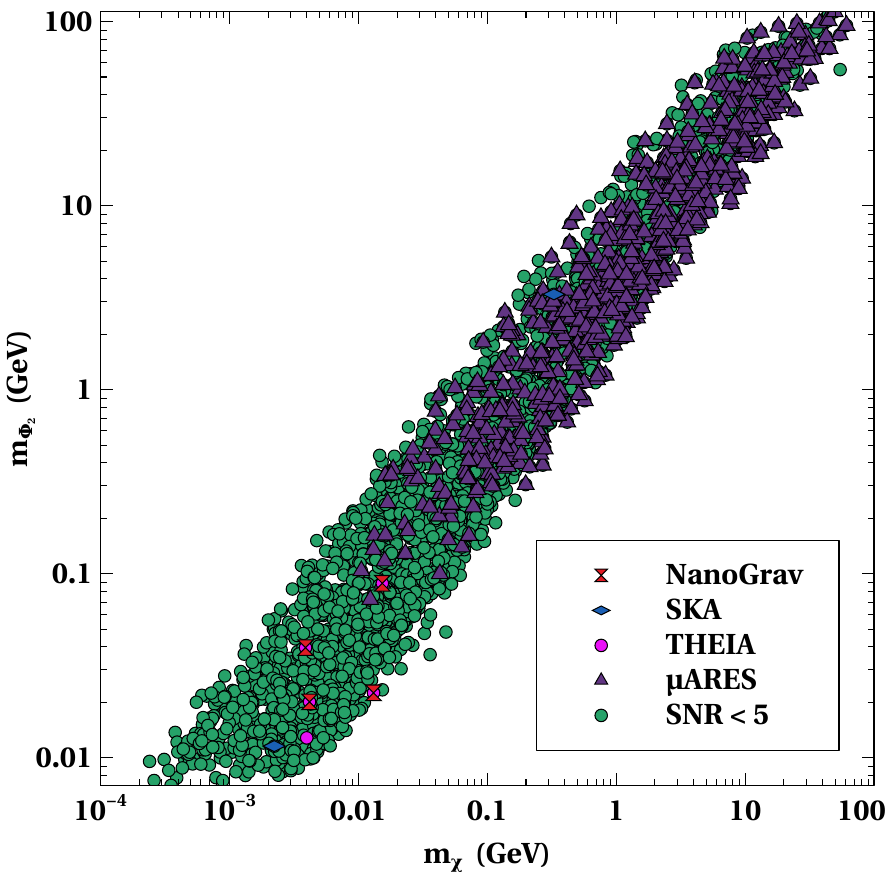}
    \caption{Parameter space in $m_\chi$-$m_{\Phi_2}$ plane with colour code indicating SNR greater than 5 for future experiments and greater than 1 for ongoing experiments.}
    \label{summary3}
\end{figure}

We show the summary in Fig. \ref{summary1} and Fig. \ref{summary3}. The left panel of Fig. \ref{summary1} shows the parameter space in effective Yukawa coupling $y_1$ versus DM mass plane for benchmark point BP1 shown in table \ref{tab1}. Different contours correspond to different values of neutrino asymmetry. The region preferred to explain the $^4{\rm He}$ anomaly is shaded in blue colour. The other experimental sensitivities and Planck 2018 bounds are also shown as different coloured shades. For a fixed value of DM mass, larger $y_1$ leads to larger neutrino asymmetry as more and more dark sector asymmetry gets transferred into neutrinos. The right panel of Fig. \ref{summary1} shows $\Delta N_{\rm eff}$ versus DM mass parameter space while different shapes of points indicating the prospect of GW discovery in terms of signal to noise ratio (SNR). The SNR 
is defined as~\cite{Schmitz:2020syl} 
\begin{equation}
\rho = \sqrt{\tau\,\int_{f_\text{min}}^{f_\text{max}}\,df\,\left[\frac{\Omega_\text{GW}(f)\,h^2}{\Omega_\text{expt}(f)\,h^2}\right]^2}\,, 
\end{equation}
with $\tau$ being the observation time for a particular detector, which we consider to be 1 yr. We set the criteria SNR $>5 (>1)$ for future (present) GW experiments to identify the parameter space. Clearly, the parameter space remains within reach of several GW experiments while at the same time offering complementary probe at future CMB experiments. As expected, lower values of DM mass requires the FOPT to occur at a lower scale to generate the finite temperature window for DM decay, pushing the corresponding peak frequencies towards the PTA ballpark. For the summary plot on the right panel of Fig. \ref{summary1} and Fig. \ref{summary3}, $v_{D}$ is varied from $10^{-2}$ GeV to $10^2$ GeV, $\lambda_{\Phi_1\Phi_2}$ is varied from 1 to 3 and $Y_{\Delta \chi}$ is varied between $5\times 10^{-4}$ and $5\times 10^{-2}$, $y_1$ is varied from 8$\times10^{-10}$ to 1$\times10^{-7}$, while $\lambda_1=\lambda_2=0.01$, and $\lambda_{12}=10^{-3}$. In Fig. \ref{summary3}, we show the zero temperature masses of dark sector particles namely DM and $\Phi_2$ satisfying all requirements and being within sensitivity of different GW experiments indicated in terms of colour code and point shape.

\begin{table*}[h!]

    \centering
    \begin{tabular}{|c|c|c|c|c|c|c|c|c|}
    \hline
      &  $v_{D}$   & $\mu_{\Phi_2}$ & $m_{\Phi_2}$    & $\lambda_{\Phi_1\Phi_2}$  & $\lambda_{1}$  & $\lambda_{2}$ & $\lambda_{\chi\Phi_1}$ & $\lambda_{12}$ \\
        
        &  (MeV)  & (MeV) & (MeV)  &    &  &  &  & \\
        \hline
     BP1 & 20 &  5 & 20.25  & 1.92 & 0.01 &  0.01 & 0.01 & $10^{-3}$ \\ 
     \hline
     BP2 & 40 &  10 & 40.49  & 1.92 &  0.01 & 0.01 & 0.01 & $10^{-3}$ \\
     \hline
     BP3 & 60 &  20 & 63.48  & 2.01 &  0.01 & 0.01 & 0.01 & $10^{-3}$ \\ 
     \hline
    \end{tabular}

    \caption{Benchmark points BP1, BP2, BP3 for different input model parameters.}
     \label{tab1}
    \end{table*}

\begin{table*}[h!]

    \centering
    \begin{tabular}{|c|c|c|c|c|c|c|c|}
    \hline
      & $T_c$  & $v_c$  & $T_n$  & $\alpha_*$ & $\beta/\mathcal{H}_* $ & $v_J$ & $T_{\rm RH}$\\
        
     & (MeV)  &  (MeV) &  (MeV) &  &  &   & (MeV)\\
        \hline
     BP1 & 6.45 & 19.00 & 2.70 & 0.12 & 40.37 & 0.79 & 2.70\\ 
     \hline
     BP2 & 12.92 & 38.00 & 5.45 & 0.12 & 46.41 & 0.79 & 5.45\\
     \hline
     BP3 & 19.87 & 57.00 & 9.04 & 0.08 & 56.93 & 0.76 & 9.04\\ 
     \hline
    \end{tabular}

    \caption{Benchmark points BP1, BP2, BP3 with other details involved in the GW spectrum calculations.}
     \label{tab1a}
    \end{table*}






\begin{table*}[h!]

    \centering
    \begin{tabular}{|c|c|c|c|c|c|c|c|c|}
    \hline
      &  $v_{D}$  & $\mu_{\Phi_2}$ & $m_{\Phi_2}$  & $\lambda_{\Phi_1\Phi_2}$  & $\lambda_{1}$  & $\lambda_{2}$ & $\lambda_{\chi\Phi_1}$ & $\lambda_{12}$ \\
        
        &  (GeV) &  (GeV) & (GeV)  &    &   &  &  &  \\
        \hline
     BP4 & 1 &  0.2 & 0.98  & 1.86 & 0.01 &  0.01 & 0.01 &  $10^{-3}$ \\ 
     \hline
     BP5 & 10 & 0.2 & 9.34   & 1.74 &  0.01 & 0.01 & 0.01 & $10^{-3}$ \\
     \hline
     BP6 & 100  & 20 & 98.36  & 1.86 &  0.01 & 0.01 & 0.01 & $10^{-3}$ \\ 
     \hline
    \end{tabular}

    \caption{Benchmark points BP4, BP5, BP6 for different input model parameters.}
     \label{tab2}
    \end{table*}

    \begin{table*}[h!]

    \centering
    \begin{tabular}{|c|c|c|c|c|c|c|c|}
    \hline
      &  $T_c$  & $v_c$  & $T_n$  & $\alpha_*$ & $\beta/\mathcal{H}_* $ & $v_J$ & $T_{\rm RH}$\\
        
        &  (GeV)  &  (GeV) &  (GeV) &   &  &   & (GeV)\\
        \hline
     BP4  & 0.32 & 0.95 & 0.13 &  0.12 & 63.64 & 0.78 & 0.13\\ 
     \hline
     BP5  & 3.14 & 9.50 & 1.12 & 0.23 & 51.01 & 0.83 & 1.12\\
     \hline
     BP6  & 32.14 & 95.00 & 12.30 & 0.15 & 33.64 & 0.80 & 12.30\\ 
     \hline
    \end{tabular}

    \caption{Benchmark points BP4, BP5, BP6 with other details involved in the GW spectrum calculations.}
     \label{tab2a}
    \end{table*}

\begin{figure}
    \centering
       \includegraphics[width=0.45\linewidth]{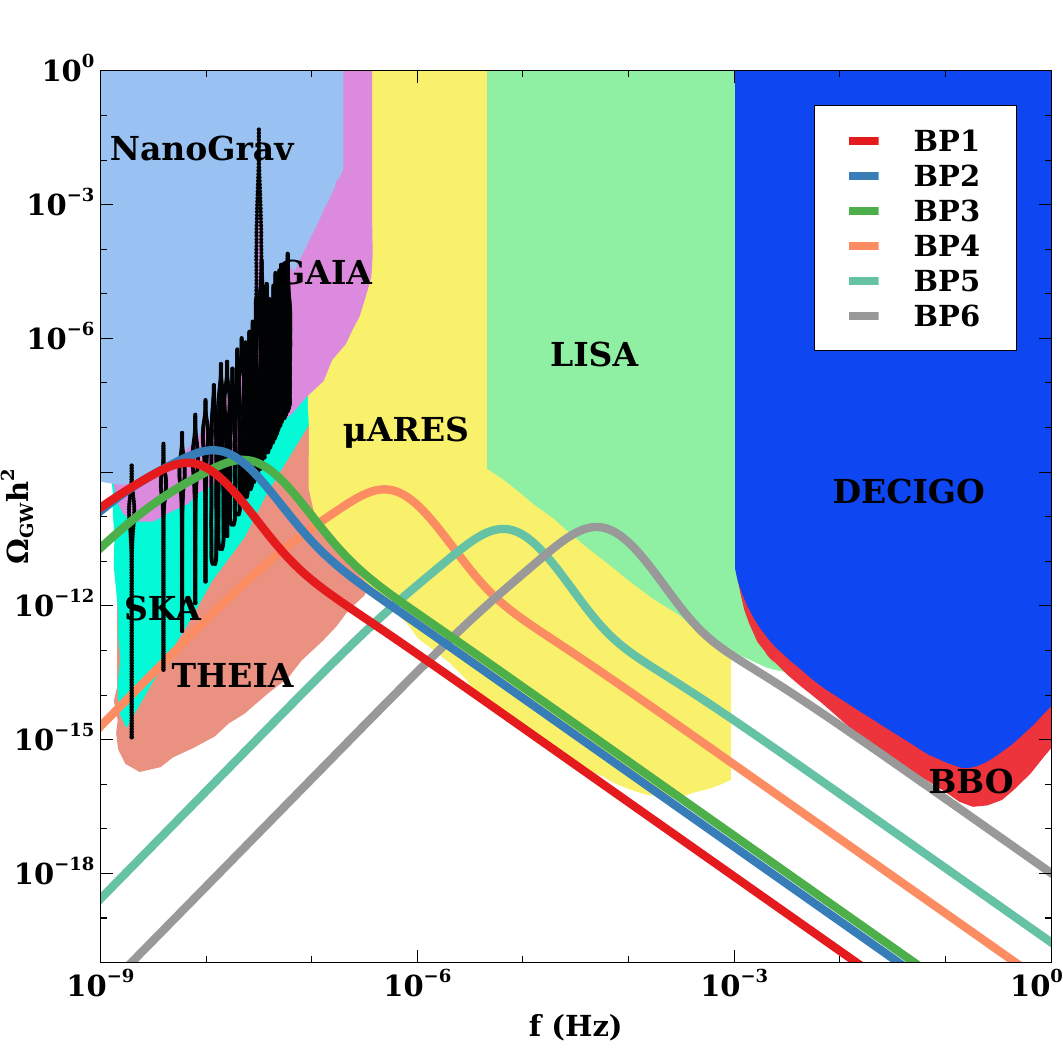}
    \caption{GW spectra for BP1 to BP6 of table \ref{tab1a}, \ref{tab2a}.}
    \label{fig:gw1}
\end{figure}

\section{Possible UV completion}
\label{sec4}
In the above discussions, we considered a low energy effective framework with explicit coupling among $\chi, \Phi_2$ and SM neutrinos. Due to the singlet nature of $\chi, \Phi_2$ under the SM gauge symmetry, such a coupling is not allowed. However, it can arise at low energy limit of a UV complete theory. The same UV completion can also explain the origin of dark sector asymmetry with possible connections to the baryon asymmetry of the Universe. There have been several proposals for asymmetric DM in the literature \cite{Nussinov:1985xr, Kaplan:2009ag, Davoudiasl:2012uw, DuttaBanik:2020vfr, Barman:2021ost, Cui:2020dly, Falkowski:2011xh, Patel:2022xyv, Biswas:2018sib,Narendra:2018vfw,Nagata:2016knk, Arina:2011cu, Arina:2012fb, Arina:2012aj, Narendra:2019cyt,Mahapatra:2023dbr,Borah:2022qln, Borah:2023qag, Borah:2024wos}. 

For simplicity, here we consider a low scale seesaw scenario based on inverse seesaw \cite{Mohapatra:1986aw, Mohapatra:1986bd, Gonzalez-Garcia:1988okv} where the desired dark fermion coupling to active neutrinos via active-sterile neutrino mixing can be realised while being consistent with light neutrino mass and sub-dominant washout of dark fermion asymmetry. Inverse seesaw scenario requires two types of sterile fermions $N, S$ with the following Lagrangian
\begin{equation}
     -\mathcal{L} \supset y_D \overline{L}\Tilde{H}N +M_S \overline{N} S+ \frac{1}{2}\,\mu_S\,\overline{S^c}S + {\rm h.c.}
 \end{equation}
Here $\mu_S$ is the only lepton number violating term which can be naturally small. After electroweak symmetry breaking, we have a Dirac mass term $M_D=\frac{y_D v}{\sqrt{2}}$ with $v$ being the VEV of the SM Higgs doublet $H$. In the limit $\mu_S \ll M_D \ll M_S$, light neutrino mass at leading order is given by
\begin{equation}
    m_\nu = M_D M^{-1}_S \mu_S (M^T_S)^{-1} M^T_D.
\end{equation}
Thus, smallness of $\mu_S$ keeps active neutrinos light while also allowing large active-sterile mixing 
\begin{equation}
    \theta^2_{\nu N} = \frac{m_\nu}{\mu_S}.
    \label{asmix}
\end{equation}
In order to generate effective $\chi-\nu$ coupling at low energy (given by Eq. \eqref{eq2}), we couple the RHN $N$ and dark sector particles $\chi, \Phi_2$ as
 \begin{equation}
     -\mathcal{L} \supset y_\chi \overline{\chi}\Phi_2 N + {\rm h.c.}
 \end{equation}
After the electroweak symmetry breaking, the heavy RHN mixes with the active neutrinos which opens up an effective interaction $y_1 \overline{\chi}\Phi_2\nu$, where $y_1 \sim y_\chi \theta_{\nu N}$ with $\theta_{\nu N}$ being the active-sterile neutrino mixing angle given by Eq. \eqref{asmix}. Given an alternate source of dark sector asymmetry, DM can carry this large asymmetry until it gets partially transferred to the neutrino sector below sphaleron decoupling temperature.

\begin{figure}
    \centering
    \includegraphics[width=0.7\linewidth]{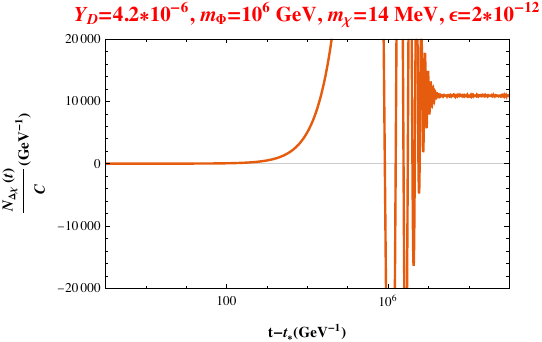}
\caption{Evolution of dark sector asymmetry with time for a benchmark choice of model parameters. The final asymmetry is of the same order as the one required to address the Helium-4 anomaly.}
    \label{fig:AD}
\end{figure}

In order to generate the large initial dark sector asymmetry, we outline a setup based on the Affleck-Dine (AD) mechanism \cite{Affleck:1984fy}. While the same mechanism can also explain the observed baryon asymmetry of the Universe, we focus only on the aspect of generating large neutrino asymmetry, which is the focus of this work. Consider an Affleck-Dine field $\Phi$ singlet under the SM symmetries but charged under a dark global $U(1)_D$ symmetry. Let the $U(1)_D$ charge of $\Phi$ be -2 while that of dark matter $\chi$ is 1.  The relevant part of the Lagrangian involving $\Phi$ is 
\begin{align}
    \mathcal{L} \supset  -(Y_D \overline{\chi^c} \chi \Phi  +{\rm h.c.})- V(\Phi),
\end{align}
where
\begin{align}
   V(\Phi)= m_{\Phi}^2 |\Phi|^2  + \lambda_{\Phi} |\Phi|^4 + (\epsilon m_{\Phi}^2 \Phi^2 + {\rm h.c.}) 
\end{align}
For $\epsilon \neq 0$, the last term in the potential $V(\Phi)$ breaks global $U(1)_D$ explicitly. While there can be other terms too breaking $U(1)_D$ explicitly, the quadratic term used here helps to obtain an analytical expression for the asymmetry \cite{Mohapatra:2021aig}. For small explicit $U(1)_D$ breaking term, we have an approximately conserved Noether's current given by
\begin{equation}
    J_\mu = iQ_\Phi [\Phi^\dagger \partial_\mu \Phi-(\partial_\mu \Phi^\dagger) \Phi],
\end{equation}
with $Q_\Phi$ being the global charge of $\Phi$. Using $\Phi=(\phi_1+i\phi_2)/\sqrt{2}$, the number density of the approximately conserved charge is 
\begin{equation}
    n_{\Delta \chi} = Q_\Phi (\dot{\phi_1} \phi_2 - \phi_1 \dot{\phi_2}).
\end{equation}
Due to this explicit violation, a non-zero lepton asymmetry is created through the cosmic evolution of $\Phi$ which gets transferred to the dark sector as $\Phi$ decays into $\chi$. For $ \Phi \gtrsim \frac{m_{\Phi}}{\sqrt{\lambda_{\Phi}}} (\equiv \Phi^*)$ (where $M_P$ denotes reduced Planck mass), the quartic term $\lambda_{\Phi} \lvert \Phi \rvert^{4}$ dominates and $\Phi\propto1/a$. Once $\Phi$ reaches $\Phi^*$, difference in the real and imaginary values of $\Phi$ creates an asymmetry in the $\Phi$ condensate which oscillates with period $T_{\rm asy}=\frac{\pi}{\epsilon m_{\Phi}}$. Now, the comoving asymmetry ($N_{\Delta \chi}(t) \equiv \left(\frac{a(t)}{a_{I}}\right)^3 n_{\Delta \chi}(t)$, with $a_{I}$ being the scale factor at the end of inflation) generated for $t>t_{*}$ can be written as \cite{Mohapatra:2021aig, Borah:2022qln} 
\begin{align}
    N_{\Delta \chi} (t)&\simeq 4 Q_\Phi \, A \, \phi_{1,I} \, \phi_{2,I} \left(\frac{\phi_I}{\Phi^*} \right) 
 \int_{t_*}^t dt' \, {\cos}(m_{1}(t'-t_*)) \, {\cos}(m_{2}(t'-t_*))  \, e^{- \Gamma_\Phi (t'-t_*)} 
\end{align}
where $\Gamma_\Phi$ indicates the total decay rate of the inflaton $\Phi$ to $\chi$. $\phi_{1,I} \, \phi_{2,I}$ indicate the initial values of the real and imaginary parts of $\Phi$ and $\phi_I=\sqrt{(\phi_{1,I})^2+(\phi_{2,I})^2}$, while $m^2_1=m^2_\Phi-2A$, $m^2_2=m^2_\Phi +2A$ with $A=\epsilon m^2_\Phi$. Asymmetry thereby created is  transferred to the dark sector through decay $\ensuremath{\Phi\rightarrow\chi\chi}$. Under the assumption of $2 A \gg \Gamma_\Phi m_\Phi$, for $t\gtrsim 1/\Gamma_{\Phi}$, the integral above simplifies to\footnote{Here, we have used the analytical expression derived in \cite{Mohapatra:2021aig} $\mathcal{I} \equiv \int_{t_*}^t dt' \, {\cos}(m_{1}(t'-t_*)) \, {\cos}(m_{2}(t'-t_*))  \, e^{- \Gamma_\Phi (t'-t_*)} \simeq \frac{\gamma}{2m_\Phi} \left ( \frac{1}{2+\gamma^2-2\sqrt{1-4\epsilon^2}}+ \frac{1}{2+\gamma^2+2\sqrt{1-4\epsilon^2}} \right) $ and assume $2\epsilon \gg \gamma$ or equivalently, $2 A \gg \Gamma_\Phi m_\Phi$ to maximize the asymmetry.}
\begin{align}
    N_{\Delta \chi} (t)&\simeq C \frac{\gamma}{8 \epsilon^2 m_{\Phi}},
    \label{eqn:NB}
\end{align}
where $\gamma=\Gamma_{\Phi}/m_{\Phi}$ and $C= 4 Q_\Phi \, A \, \phi_{1,I} \, \phi_{2,I} \left(\frac{\phi_I}{\Phi^*} \right)$. Also, we have $\gamma \ll 1$ for narrow width and $\epsilon \ll 1$ for soft breaking of global symmetry. In Fig. \ref{fig:AD}, we show the evolution of the comoving asymmetry $N_{\Delta \chi} (t)$. The asymmetry initially rises from zero and then oscillates until $t\gtrsim 1/\Gamma_{\Phi}$, when its amplitude exponentially damps to reach the constant value given by Eq. \eqref{eqn:NB}. The final constant value of the asymmetry is consistent with the neutrino asymmetry required to solve the recently reported Helium-4 anomaly.

\begin{figure}
    \centering
    \includegraphics[width=0.7\linewidth]{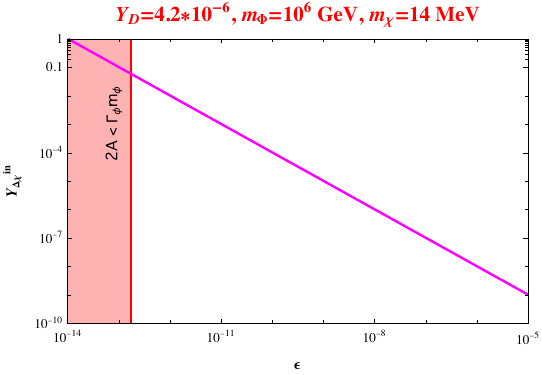}
\caption{Comoving dark sector asymmetry as a function of $\epsilon$ for one benchmark choice of model parameters. The analytical expression for asymmetry is not valid in the pink shaded region requiring numerical estimates.}
    \label{fig:AD1}
\end{figure}

If the same decay of the AD field $\Phi$ to dark matter also reheats the Universe to a temperature $T_{\rm rh}$, then we can relate its decay width $\Gamma_\Phi$ to the reheat temperature $T_{\rm rh}\simeq\sqrt{\Gamma_{\Phi}M_{\rm P}}$. The lepton asymmetry initially rises from zero and then oscillates until $t\gtrsim 1/\Gamma_{\Phi}$, when its amplitude exponentially damps to reach the constant value given by \cite{Lloyd-Stubbs:2020sed,Mohapatra:2021aig, Borah:2022qln}

\begin{equation}
Y^{\rm in}_{\Delta \chi} = \frac{(n_{\chi}-n_{\bar{\chi}})}{s}\Big|_{T_{\rm rh}} = \frac{N_{\Delta \chi}(T_{\rm rh})}{s(T_{\rm rh})}\left(\frac{a_{\rm I}}{a_{\rm rh}}\right)^{3} = \frac{N_{\Delta \chi}(T_{\rm rh})}{s(T_{\rm rh})} \left(\frac{a_{\rm I}}{a_{*}}\right)^{3} \left(\frac{a_{*}}{a_{\rm rh}}\right)^{3}.
\end{equation}
Using $a \propto 1/\Phi$ for $a < a_*{}$ and $a \propto t^{2/3} \propto \mathcal{H}^{-2/3}$ for $a_{*}<a<a_{\rm rh}$, we obtain 
\begin{equation}
    Y^{\rm in}_{\Delta \chi} \simeq \frac{N_{\Delta \chi}(T_{\rm rh})}{s(T_{\rm rh})} \left(\frac{\Phi_{*}}{\phi_{I}}\right)^{3} \left(\frac{\mathcal{H}_{\rm rh}}{\mathcal{H}_{*}}\right)^2.
\end{equation}
Using $\mathcal{H}^2_{\rm rh} = \frac{\pi^2}{90} g_{*} \frac{T^4_{\rm rh}}{M^2_{P}}$, $\mathcal{H}^2_{*} = \frac{m^2_{\Phi}\Phi_{*}^2}{6 M^2_{P}}$ and the expression for $N_{\Delta \chi}(T_{\rm rh})$ given in Eq. \eqref{eqn:NB}, we obtain
\begin{equation}
Y^{\rm in}_{\Delta \chi} \simeq \frac{3}{4}\sqrt{\frac{\pi^2}{90}g_{*}} Q_{\Phi} \phi_{1,I} \phi_{2,I}\left(\frac{1}{\phi_{I}}\right)^2 \frac{T_{\rm rh}^{3}}{\epsilon m_{\Phi}^{2}M_{\rm P}} \simeq \frac{3}{8}\sqrt{\frac{\pi^2}{90}g_{*}} Q_{\Phi} \frac{T_{\rm rh}^{3}}{\epsilon m_{\Phi}^{2}M_{\rm P}} \sin(2\theta_{\phi}).\label{eq:vis}
\end{equation}
Here $\theta_{\phi}$ is the phase of $\Phi$, which we can consider to be of $\mathcal{O}(1)$.
 Now, the presence of lepton number violating interaction given by $\epsilon$ can lead to the washout of the generated asymmetry. This can happen through scatterings violating $U(1)_D$ charge by 4 units: $\chi \chi \leftrightarrow \overline{\chi}~\overline{\chi}$, mediated by $\Phi$ exchange and the $\epsilon$ term. If the decoupling temperature of such process is higher than the reheat temperature $T_{\rm rh}$, the washout effect would be absent. This leads to the following condition
\begin{equation}
\Gamma_{\rm wo} (\Delta L=4) \equiv T_{\rm rh}^{3}\frac{Y_{D}^{4}\epsilon^2T_{\rm rh}^2}{64 \pi m_{\Phi}^{4}}\lesssim \mathcal{H}(T_{\rm rh}) \equiv \sqrt{\frac{\pi^2}{90}g_*} \frac{T_{\rm rh}^{2}}{M_{\rm P}}.\label{eq:washout}
\end{equation}
Additionally, demanding the AD field to be not part of the thermal bath requires $T_{\rm rh} < m_\Phi$ which ensures washouts due to inverse decays being sub-dominant. Fig. \ref{fig:AD1} shows comoving dark asymmetry as a function of $\epsilon$ for a benchmark choice of parameters. The pink shaded region corresponds to the region where analytical approximation of asymmetry given by Eq. \eqref{eqn:NB} does not apply. The entire parameter space shown in Fig. \ref{fig:AD1} correspond to sub-dominant washouts mediated by $\Phi$ or washouts due to inverse decay into $\Phi$.

Another washout of dark asymmetry can arise due to the coupling of $\chi$ with Majorana heavy fermions $N_i$. One can have $\Delta L=2$ washout process like $\chi \Phi^\dagger_2 \leftrightarrow \overline{\chi}~\Phi_2$ mediated by $N_i$. Demanding this process to be inefficient at $T=T_{\rm rh}$ leads to
\begin{equation}
\Gamma_{\rm wo} (\Delta L=2) \equiv T_{\rm rh}^{3}\frac{y_{\chi}^{4} \mu^2_S}{64 \pi M_{S}^{4}} = T_{\rm rh}^{3}\frac{y_{1}^{4} \mu^2_S}{64 \pi (\theta_{\nu N} M_{S})^{4}} \lesssim \mathcal{H} (T_{\rm rh}) \equiv \sqrt{\frac{\pi^2}{90}g_*} \frac{T_{\rm rh}^{2}}{M_{\rm P}},\label{eq:washout2}
\end{equation}
where we use $y_{1}=y_{\chi}\theta_{\nu N}$. Using active-sterile mixing given by Eq. \eqref{asmix}, this leads to the following upper bound
\begin{equation}
 T_{\rm rh} < \sqrt{\frac{\pi^2}{90}g_*} \frac{64\pi}{y^4_1} \frac{m^2_\nu}{M_{\rm P}} \left ( \frac{M_S}{\mu_S} \right)^4.
\end{equation}
For $y_1 \sim 10^{-9}$ and $m_\nu < 0.1$ eV, $T_{\rm rh}$ can be as large as $10^{16}$ GeV for $M_S/\mu_S \sim 10^4$. For larger $M_S/\mu_S$ which is generic in inverse seesaw scenarios, this bound can be satisfied trivially.

In order not to overproduce baryon asymmetry of Universe, it is also required to ensure that the large dark fermion asymmetry does not get transferred to SM leptons via scatterings like $\chi \Phi^\dagger_2 \rightarrow l H$ mediated by $N$. Keeping this $\Delta L=0$ conversion process out-of-equilibrium at $T=T_{\rm rh}$ leads to
\begin{equation}
\Gamma_{\rm transfer} \equiv T_{\rm rh}^{3}\frac{y^2_D y_{\chi}^{2}}{64 \pi M_{S}^{2}}=T_{\rm rh}^{3}\frac{y^2_1 }{32 \pi v^{2}}\lesssim \mathcal{H}(T_{\rm rh}) \equiv \sqrt{\frac{\pi^2}{90}g_*} \frac{T_{\rm rh}^{2}}{M_{\rm P}},\label{eq:washout3}
\end{equation}
where we use  $y_{\chi}=\frac{y_{1}}{\theta_{\nu N}}=\frac{\sqrt{2}\,y_{1}\, M_{S}}{y_{D}\, v}$.

\begin{figure}
    \centering
    \includegraphics[width=0.9\linewidth]{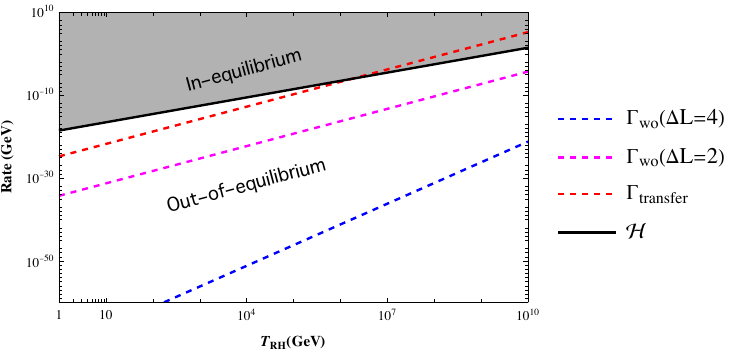}
\caption{Comparison of asymmetry washout rates $\Gamma_{\rm wo} (\Delta \rm L =4)$, $\Gamma_{\rm wo} (\Delta \rm L =2)$ and asymmetry transfer rate $\Gamma_{\rm transfer}$ with Hubble expansion rate $\mathcal{H}$ as a function of reheating temperature. The blue, magenta and red color dashed lines denote the interaction rate given in Eq. \eqref{eq:washout}, \eqref{eq:washout2} and \eqref{eq:washout3} whereas black solid line denotes the corresponding Hubble expansion rate.}
    \label{fig:AD2}
\end{figure}

To verify the out-of-equilibrium conditions given by Eqs. \eqref{eq:washout}, \eqref{eq:washout2} and \eqref{eq:washout3} for the benchmark plots shown in
Fig. \ref{fig:AD} and \ref{fig:AD1}, we plot the Hubble expansion rate and interaction rates for washout and transfer processes as a function of reheating temperature in Fig. \ref{fig:AD2}. The region below the black solid line denotes the out-of equilibrium regime. The blue, magenta and red colored dashed lines denote the interaction rate given in Eqs. \eqref{eq:washout}, \eqref{eq:washout2} and \eqref{eq:washout3} respectively. The rest of the parameters are kept fixed at $Y_{D}=4.2\times 10^{-6}, m_{\Phi}=10^{6}$ GeV, $m_{\chi}=14$ MeV and $\epsilon = 2\times 10^{-12}$ (same values as in Fig. \ref{fig:AD} and Fig. \ref{fig:AD1}). In addition, we use $y_{1}=10^{-9}$ which is consistent with left plot of Fig. \ref{summary1}. For the condition in Eq. \eqref{eq:washout2}, we consider the ratio $\frac{M_{S}}{\mu_{S}}$ to be $10^{4}$. With these, the maximum reheating temperature to keep all the washout and transfer processes out-of-equilibrium is found to be $\sim 10^{6}$ GeV.

\section{Conclusion}
\label{sec5}
We have proposed a mechanism to produce large neutrino asymmetry at low scale from dark matter decay. While dark matter is cosmologically stable, it can decay in the early Universe due to finite temperature effects and if DM is asymmetric like the visible matter, its decay into light neutrinos can transfer some of the dark sector asymmetries into neutrinos. The decay of DM is enabled in the vicinity of a first order phase transition which leads to sharp change in one of the dark sector particle's mass. The large neutrino asymmetry can have interesting cosmological consequences like altering light nuclei abundance which can solve the recently reported Helium-4 anomaly. Such asymmetry can also lead to observable $\Delta N_{\rm eff}$ at future CMB experiments. In order to be in agreement with the observed baryon asymmetry, such decay of DM and hence the FOPT is required to occur below the electroweak scale. This keeps the corresponding gravitational wave peak frequencies in nHz-mHz ballpark. Due to the presence of new physics below the electroweak scale, the model can have other detection prospects the details of which we do not discuss here. We briefly comment upon interesting direct detection prospects of DM via Higgs portal interactions in Appendix \ref{appen2}. While we have not explored the possibility of explaining the observed baryon asymmetry within the same setup, it can be generated independently from heavy right handed neutrino decay \cite{Gu:2010xc}. Given the upper bound on reheating temperature $T_{\rm rh} < 10^{6}$ GeV from the requirement of keeping washout and asymmetry transferring scattering processes out of equilibrium, the leptogenesis is likely to occur via resonant enhancement or non-thermally. We leave such possible extensions of the framework to future studies.

\acknowledgments
The work of D.B. is supported by the Science and Engineering Research Board (SERB), Government of India grants MTR/2022/000575 and CRG/2022/000603. The work of N.D. is supported by the Ministry of Education, Government of India via the Prime Minister's Research Fellowship (PMRF) December 2021 scheme. 

\appendix

\section{Details of FOPT and Gravitational Waves}
\label{appen1}
The dark sector went through a first order phase transition driven by scalar $\Phi_1$ and the tree level scalar potential can be written as
\begin{align}
    V(\Phi_1,\Phi_2) & =\frac{\lambda_1}{4} \left (\Phi_1^2-\frac{v_D^2}{2} \right)^2 + \mu_{\Phi_2}^2|\Phi_2|^2 + \lambda_{2}|\Phi_2|^4 + \frac{\lambda_{\Phi_1\Phi_2}}{2}\Phi_1^2 |\Phi_2|^2.
\end{align}
Here, $v_D$ is the vacuum expectation value (VEV) of the singlet scalar $\Phi_1(\equiv \phi_1)$. 
The Coleman-Weinberg (CW) potential \cite{Coleman:1973jx} can be written as
\begin{equation}
V_{\rm CW}(\phi_1)=\frac{1}{ (8\pi)^2}\sum_{i=\phi_1,\Phi_{2},\chi}n_i M_i^4(\phi_1)\left \{\log{ \left ( \frac{M_i^2(\phi_1)}{v_D^2} \right )}-C_i\right \}
\end{equation}
where, $n_{\phi_1}=1$, $n_{\Phi_{2}}=2$, $n_\chi=4$
and $C_{\phi_1,\Phi_{2},\chi}=\frac{3}{2}$. The physical field-dependent masses of particles coupled to $\Phi_1$ are
\begin{align*}
M_{\phi_{2}}^2(\phi_1)=\mu_{\Phi_2}^2+\frac{\lambda_{\Phi_1\Phi_2}}{2}\phi_1^2, \hspace{0.1 cm} 
M_\chi^2(\phi_1)=(m_0+y_{\chi\Phi_1}\phi_1)^2.
\end{align*}
Now, the thermal contributions \cite{Dolan:1973qd,Quiros:1999jp} can be expressed as
\begin{equation}
V_T(\phi_1,T)=\sum_{i=\phi_1,\Phi_{2},\chi}\frac{n_iT^4}{2\pi^2}J_B \left(\frac{M_i^2(\phi_1)}{T^2}\right) 
\end{equation}
where
$$ J_B(x)=\int_{0}^{\infty}dy\, y^2 \log[1-e^{-\sqrt{y^2+x^2}}].$$
In the thermal contribution, the Daisy corrections \cite{Fendley:1987ef,Parwani:1991gq,Arnold:1992rz} are also added  to improve perturbative expansion following Arnold-Espinosa method \cite{Arnold:1992rz} where $V_{\rm thermal}(\phi_1,T)=V_T(\phi_1,T) + V_{\rm daisy}(\phi_1,T)$. The Daisy contribution can be written as
\begin{equation}
V_{\rm daisy}(\phi_1,T)=-\frac{T}{2\pi^2}\sum_{i=\phi_1,\Phi_{2}} n_i[M_i^3(\phi_1,T)-M_i^3(\phi_1)]
\end{equation}
where, $M_i^2(\phi_1,T)$=$M_i^2(\phi_1)$ + $\Pi_i(T)$ and the relevant thermal masses are
\begin{align*}
    &M_{\phi_1}^2(\phi_1,T)= M_{\phi_1}^2(\phi_1)+\left(\frac{\lambda_{1}}{4} +\frac{\lambda_{\Phi_1 \Phi_2}}{6} \right)T^2, \hspace{0.4 cm}\\
     &M_{\Phi_{2}}^2(\phi_1,T)= M_{\Phi_{2}}^2(\phi_1)+\left(\frac{\lambda_{2}}{3}+\frac{\lambda_{\Phi_1 \Phi_2}}{12}  \right)T^2.
\end{align*}
Hence, the effective potential at finite temperature can be written as 
\begin{equation}
    V_{\rm eff}(\phi_1,T)=V_{\rm tree}(\phi_1) + V_{\rm CW}(\phi_1) + V_{\rm thermal} (\phi_1,T) .\label{eq:Veff}
\end{equation}

The above effective potential shows different profiles depending on the temperature. When the temperature reaches a critical point, the potential develops two identical minima separated by a barrier. Below the critical temperature, the non-zero minimum becomes the true vacuum, while the zero minimum becomes the false vacuum. Due to the presence of the barrier, the Universe remains in the false vacuum, and then transitions from the false vacuum to the true vacuum by tunneling through the barrier. The rate of tunneling per unit time per unit volume can be estimated as 
$\Gamma (T) = \mathcal{A}(T) e^{-S_3(T)/T},$
where $\mathcal{A}(T)\sim T^4$ and $S_3$ is Euclidean action. 
%
Here, the Hubble parameter is given by ${\bf \mathcal{H}}(T)\simeq 1.66\sqrt{g_*}T^2/M_{\rm P}$ with $g_*$ being the dof of the radiation component. The energy difference between the true and the false vacuum is $\Delta V_{\rm eff} \equiv V_{\rm eff}(\phi_{\rm false},T)- V_{\rm eff}(\phi_{\rm true},T).$ The amount of vacuum energy released during FOPT can parameterized as $\alpha_* =\frac{\epsilon_*}{\rho_{\rm rad}}$, where radiation energy density of the Universe, $\rho_{\rm rad}= g_*\pi^2 T^4/30 $ and 
$ \epsilon_* = \left[\Delta V_{\rm eff} - \frac{T}{4} \frac{\partial \Delta V_{\rm eff}}{\partial T}\right]_{T=T_n}.$ The inverse duration of FOPT is defined as $\frac{\beta}{{\mathcal{ H}}(T)_*} \simeq T\frac{d}{dT} \left(\frac{S_3}{T} \right)|_{T=T_n}.$ In this work to calculate the Euclidean action, we fitted the effective potential to a generic potential for which action calculations are described in \cite{Adams:1993zs}. This prescription of action estimation is described with details in \cite{Borah:2022cdx}.

As the FOPT progresses, bubbles of true vacuum form and expand to cover the whole Universe. Then these bubbles start to collide with each other and generate stochastic gravitational waves. The gravitational waves are produced from three primary sources: bubble collisions, sound waves in the plasma, and turbulence due to MHD of the plasma. Now, considering these three contributions to GW production, the corresponding GW power spectrum can be written as \cite{Cai:2017tmh}
\begin{align}
    \Omega_{\rm GW}(f) &= \Omega_\phi(f) + \Omega_{\rm sw}(f) + \Omega_{\rm turb}(f).
\end{align}
 Considering bubble collision as one of the source, the spectrum can be written as 	\cite{Lewicki:2022pdb, Athron:2023xlk, Caprini:2024hue}
 \begin{equation}
\Omega_\phi h^2 = 1.67 \times 10^{-5} \left ( \frac{100}{g_*} \right)^{1/3} \left(\frac{\mathcal{H}_*}{\beta}\right)^2 \left(\frac{\kappa_\phi \alpha_*}{1+\alpha_*}\right)^2 \frac{A (a+b)^c}{\left[ b (f/f^\phi_{\rm peak})^{-a/c}+ a(f/f^\phi_{\rm peak})^{b/c} \right]^c} ,
\end{equation}
where, $a=2.41, b=2.42, c=4.08$ and A=5.13$\times10^{-2}$ and the peak frequency being 
	\begin{equation}
 f^\phi_{\rm peak} = 2.02 \times 10^{-6} {\rm Hz} \left ( \frac{g_*}{100} \right)^{1/6} \left ( \frac{T_n}{100 \; {\rm GeV}} \right )  \left(\frac{\beta}{\mathcal{H}_*}\right).
	\end{equation}
 During bubble collision, the energy transfer can be parameterised by the  efficiency factor $\kappa_\phi$, \cite{Kamionkowski:1993fg}
 \begin{equation}
     \kappa_\phi =\frac{1}{1+0.715\alpha_*}\left( 0.715 \alpha_* + \frac{4}{27} \sqrt{\frac{3\alpha_*}{2}} \right).
 \end{equation}
The Jouguet velocity for bubble can be written as  $v_J = \frac{1/\sqrt{3} + \sqrt{\alpha^2_* + 2\alpha_*/3}}{1+\alpha_*}$ \cite{Espinosa:2010hh}  and the bubble wall velocity can be estimated as  \cite{Lewicki:2021pgr}
\begin{equation}
    v_w = 
    \begin{cases}
    \sqrt{\frac{\Delta V_{\rm eff}}{\alpha_* \rho_{\rm rad}}} & \text{if} \,\, \sqrt{\frac{\Delta V_{\rm eff}}{\alpha_* \rho_{\rm rad}}} < v_J \\
    1 & \text{if}  \,\, \sqrt{\frac{\Delta V_{\rm eff}}{\alpha_* \rho_{\rm rad}}} \geq v_J. \\
    \end{cases}
\end{equation}

	The GW spectrum generated from the sound wave in the plasma can be written as \cite{Athron:2023xlk} 
	\begin{equation}
		\Omega_{\rm sw}h^2 = 2.59\times 10^{-6} \left(\frac{100}{g_*}\right)^{1/3}\left(\frac{\mathcal{H}_*}{\beta}\right) \left( \frac{\kappa_{\rm sw} \alpha_*}{1+\alpha_*}\right)^2 v_w\frac{7^{3.5}(f/f^{\rm sw}_{\rm peak})^3}{(4+3(f/f^{\rm sw}_{\rm peak})^2)^{3.5}} \Upsilon.
	\end{equation}
	The corresponding peak frequency is given by
	\begin{equation}
		f^{\rm sw}_{\rm peak}=8.9\times10^{-6}{\rm Hz} \left ( \frac{g_*}{100} \right)^{1/6} \frac{1}{v_w} \left ( \frac{T_n}{100 \; {\rm GeV}} \right )  \left(\frac{\beta}{\mathcal{H}_*}\right)(\frac{z_p}{10}).
	\end{equation}
 The efficiency factor for sound waves  is \cite{Espinosa:2010hh}
 \begin{equation}
     \kappa_{\rm sw} =\frac{\alpha_*}{0.73+0.083 \sqrt{\alpha_*}+ \alpha_*}.
 \end{equation}
 Here, $\Upsilon=1-\frac{1}{\sqrt{1+2\tau_{sw}\mathcal{H}_*}}$ is a suppression factor \cite{Guo:2020grp} where $\tau_{\rm sw}\sim R_*/\bar{U}_f$, mean bubble separation, $R_*=(8\pi)^{1/3}v_w \beta^{-1}$ and rms fluid velocity, $\bar{U}_f=\sqrt{3\kappa_{sw}\alpha_*/4(1+\alpha_*)}$; $z_p\sim 10$.
The spectrum generated by the turbulence in the plasma is given by \cite{Caprini:2015zlo, Athron:2023xlk, Caprini:2024hue}
	\begin{equation}
		\Omega_{\rm turb}h^2 = 3.35\times 10^{-4} \left(\frac{100}{g_*}\right)^{1/3}\left(\frac{\mathcal{H}_*}{\beta}\right) \left( \frac{\kappa_{\rm turb} \alpha_*}{1+\alpha_*}\right)^{1.5}v_w \frac{(f/f^{\rm turb}_{\rm peak})^3}{(1+f/f^{\rm turb}_{\rm peak})^{3.6}(1+8\pi f/h_*)},
	\end{equation}
   with the peak frequency being \cite{Caprini:2015zlo}
	\begin{equation}
		f^{\rm turb}_{\rm peak}=2.7\times10^{-5}{\rm Hz} \left ( \frac{g_*}{100} \right)^{1/6} \frac{1}{v_w} \left ( \frac{T_n}{100 \; {\rm GeV}} \right )  \left(\frac{\beta}{\mathcal{H}_*}\right).
	\end{equation}
 The efficiency factor for turbulence is $\kappa_{\rm turb} \simeq 0.1 \kappa_{\rm sw}$ \cite{Caprini:2015zlo} and 
	\begin{equation}
		h_*=1.65\times10^{-5}{\rm Hz} \left ( \frac{g_*}{100} \right)^{1/6} \left ( \frac{T_n}{100 \; {\rm GeV}} \right ).
	\end{equation}

\begin{figure}
    \centering
    \includegraphics[width=0.45\linewidth]{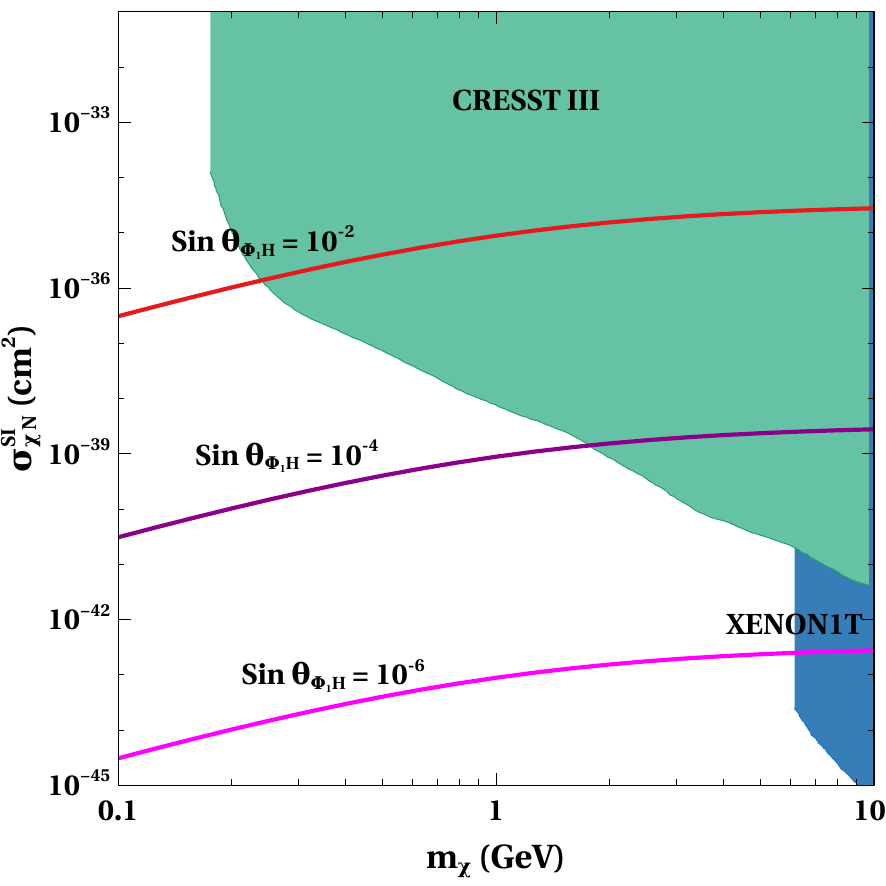}
    \includegraphics[width=0.45\linewidth]{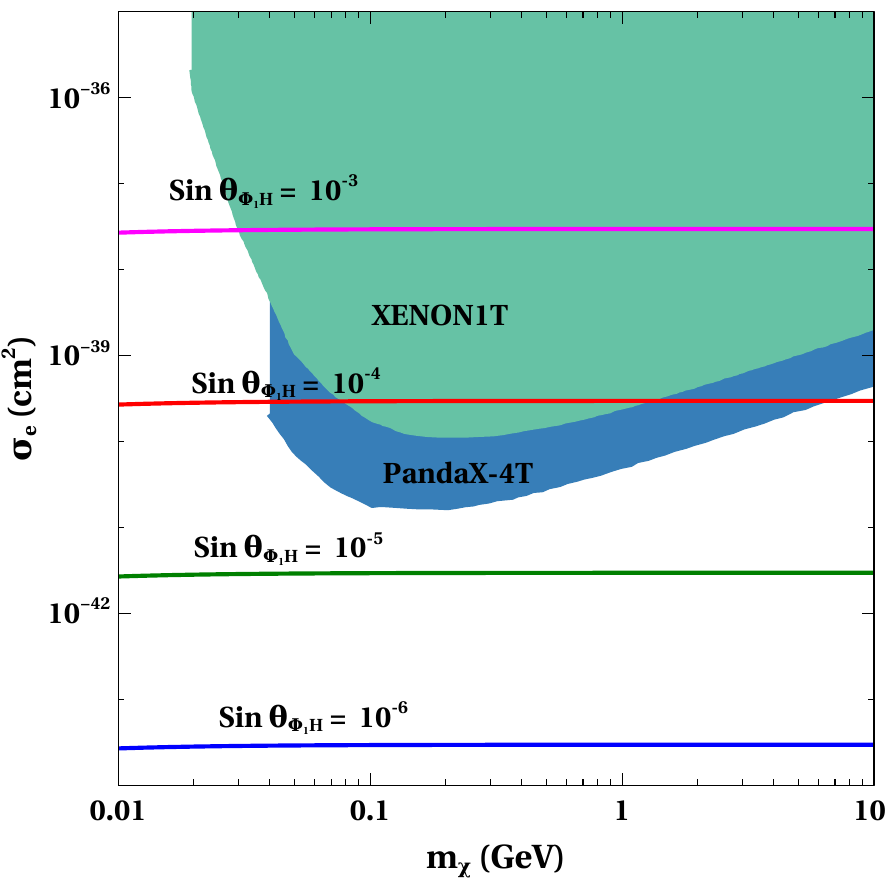}
    \caption{Left panel: Variation of spin-independent DM-nucleon elastic scattering cross section with DM mass for different choices of the singlet-Higgs mixing parameter. Right panel: Same as the left panel but for DM-electron scattering. For both the panels, $m_{\Phi_1}=10$ MeV and $\lambda_{\chi\Phi_1}$=0.01.}
    \label{fig:DD}
\end{figure}

\section{Direct detection of light DM}
\label{appen2}


DM $\chi$, can interact with nucleon due to Higgs portal interactions facilitated by the mixing between $\Phi_1$ and the SM Higgs $h$, parameterized by $\sin{\theta_{{\Phi_1}h}}$. The spin independent DM-nucleon cross section for a target nucleus with atomic number $Z$ and mass number $A$ can be written as \cite{Adhikary:2024btd}
\begin{equation}
    \sigma_{\chi N}^{SI} = \frac{\mu^2_{{\chi}N}}{4 \pi A^2} \left(Z f_{p} + (A-Z) f_{n}\right)^2,
\end{equation}
where $\mu_{{\chi}N}=\frac{m_\chi m_N}{(m_\chi+m_N)}$ is the reduced mass of DM-nucleon system and the interaction strengths $f_{p}$, $f_{n}$ of proton and neutron with DM are given as 
\begin{equation}
    f_{p,n} =  \sum_{q= u,d,s} f^{p, n}_{T_{q}}  \alpha_{q}  \frac{m_{p,n}}{m_{q}} + \sum_{q=c,t,b} f^{p,n}_{TG} \alpha_{q} \frac{m_{p,n}}{m_{q}}, 
\end{equation}
where $f^{p}_{T_{u}}= 0.018$, $f^{p}_{T_{d}}= 0.027$, $f^{p}_{T_{s}}= 0.037$, $f^{n}_{T_{u}}= 0.013$, $f^{n}_{T_{d}}= 0.040$, $f^{n}_{T_{s}}= 0.037$ and $f^{p}_{TG} = 1 - f^{p}_{T_{u}} - f^{p}_{T_{d}} - f^{p}_{T_{s}}$,  $f^{n}_{TG} = 1 - f^{n}_{T_{u}} - f^{n}_{T_{d}} - f^{n}_{T_{s}}$ \cite{Ellis:2018dmb}. 
The $\alpha_{q}$ is defined by 
\begin{equation}
    \alpha_{q} = \lambda_{\chi \Phi_1 } \sin{\theta}_{\Phi_1  h} \frac{m_{q}}{v} \left( \frac{1}{m^2_{\Phi_1}} - \frac{1}{m^2_{h}} \right),
\end{equation}
with $m_h\approx 125$ GeV being the mass of SM Higgs.


Moreover, the dark matter electron scattering in direct detection experiments XENON1T \cite{XENON:2019gfn}, PandaX-4T \cite{PandaX:2022xqx} can also provide bound on DM mass. The DM-electron cross section is given by \cite{Adhikary:2024btd, Essig:2015cda,Prabhu:2022dtm}
\begin{equation}
    \sigma_{e} \approx \frac{16 \pi \mu_{{\chi}e} \alpha \alpha_{\chi \Phi_1} \sin^{2}{\theta_{{\Phi_1}h}}} {m^4_{\Phi_1}} |F(q)|^{2},
\end{equation}
with $\alpha_{\chi \Phi_1} = \lambda^2_{\chi \Phi_1}/ 4\pi$ and $\mu_{\chi e} = \frac{m_{\chi}m_{e}}{m_{\chi}+m_{e}}$ refers the reduced mass of DM-electron system. $\alpha$ denotes the fine structure constant, and $F(q) = 1$ for a massive mediator. The variations of DM-nucleon and DM-electron scattering cross section with DM mass in the range of our interest are shown in Fig. \ref{fig:DD}.

\begin{figure}[h]
    \centering
    \includegraphics[width=0.45\linewidth]{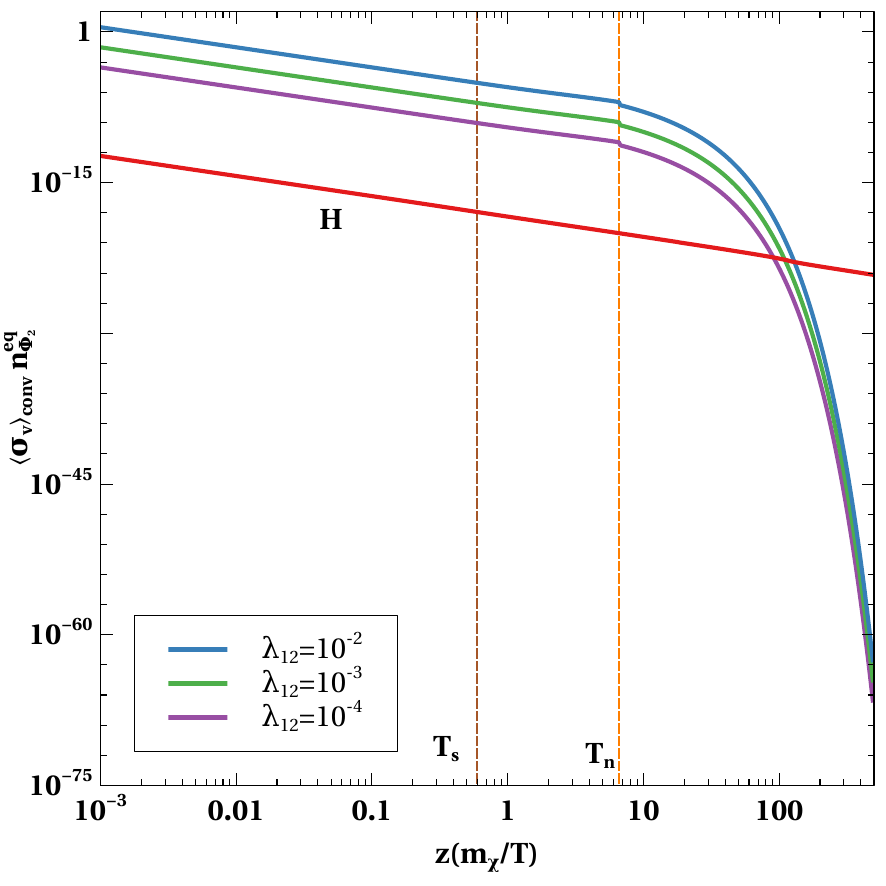}
    \caption{Comparison of conversion rate of $\Phi_2 \rightarrow \Phi^\dagger_2$  for different value of $\lambda_{12}$ and other parameters same as BP1.}
    \label{fig:rate}
\end{figure}

\color{black}

\section{Efficient conversion of $\Phi_2 \rightarrow \Phi^\dagger_2$}
\label{appen3}
The decay of asymmetric $\chi$ can also create asymmetry in $\Phi_2$, but due to the scalar coupling $\lambda_{12} (\Phi_2 \Phi_2 \Phi_1 \Phi_1+{\rm h.c.})$, efficient conversion of $\Phi_2 \rightarrow \Phi^\dagger_2$ can happen before and during FOPT, effectively washing out the asymmetry stored $\Phi_2$. The cross-section for this process is $\sigma_{\rm conv}=\lambda_{12}^2/(16\pi s)$. The corresponding thermal averaged cross section \cite{Cannoni:2015wba} can be written as
\begin{equation}
    \langle \sigma v\rangle_{\rm conv}=\frac{1}{8T m_{\Phi_1}^2m_{\Phi_2}^2K_2(m_{\Phi_1}/T)K_2(m_{\Phi_2}/T)}\int_{(m_{\Phi_1}+m_{\Phi_2})^2}^\infty ds \frac{\lambda(s,m_{\Phi_1},m_{\Phi_2})}{\sqrt{s}}K_1(\frac{\sqrt{s}}{T})\sigma_{\rm conv}
\end{equation}
where, the Mandelstam triangular function $\lambda(s,m_1,m_2)=(s-(m_1+m_2)^2)(s-(m_1-m_2)^2)$.
Fig. \ref{fig:rate} shows the comparison of this conversion rate with Hubble for different value of $\lambda_{12}$, and other parameter same as benchmark point BP1 in table \ref{tab1}. Clearly, the process remains in equilibrium around the phase transition epoch, efficiently washing out the asymmetry stored in $\Phi_2$.

\bibliographystyle{JHEP}
\bibliography{arxiv_v2.bbl}

\end{document}